\documentclass[prx, aps, 10pt, twocolumn, a4paper, superscriptaddress, longbibliography]{revtex4-1}

\usepackage{amsmath}
\usepackage{amssymb}
\usepackage{mathtools}
\usepackage{physics}
\usepackage{textcomp}
\usepackage{upgreek}
\usepackage{units}
\usepackage[pdftex]{graphicx}
\usepackage{microtype}
\usepackage{enumitem}
\usepackage{xcolor}
\usepackage{gnuplottex}    
\usepackage{multirow}   
\usepackage{hyperref}
\usepackage{placeins}
\usepackage{siunitx}
\usepackage{csquotes}
\usepackage{booktabs}
\usepackage{sidecap}

\usepackage{txfonts}

\renewcommand{\vec}[1]{\ensuremath{\boldsymbol{#1}}}

\newcommand{\mub}{\mu_\mathrm{B}}

\usepackage{pifont}
\newcommand{\iu}{\mathrm{i}}

\DeclareMathOperator{\diag}{\mathop{\mathrm{diag}}}

\newcommand{\adj}[1]{#1^\dagger}

\newcommand{\conj}[1]{#1^*}

\begin{document}

\title{
Origin of the Magnetic Spin Hall Effect: Spin Current Vorticity in the Fermi Sea
}

\author{Alexander Mook}
\affiliation{Institut f\"ur Physik, Martin-Luther-Universit\"at Halle-Wittenberg, D-06099 Halle (Saale), Germany}
\affiliation{Department of Physics, University of Basel, Klingelbergstrasse 82, CH-4056 Basel, Switzerland}

\author{Robin R. Neumann}
\affiliation{Institut f\"ur Physik, Martin-Luther-Universit\"at Halle-Wittenberg, D-06099 Halle (Saale), Germany}

\author{Annika Johansson}
\affiliation{Institut f\"ur Physik, Martin-Luther-Universit\"at Halle-Wittenberg, D-06099 Halle (Saale), Germany}

\author{J\"urgen Henk}
\affiliation{Institut f\"ur Physik, Martin-Luther-Universit\"at Halle-Wittenberg, D-06099 Halle (Saale), Germany}

\author{Ingrid Mertig}
\affiliation{Institut f\"ur Physik, Martin-Luther-Universit\"at Halle-Wittenberg, D-06099 Halle (Saale), Germany}
\affiliation{Max-Planck-Institut f\"ur Mikrostrukturphysik, D-06120 Halle (Saale), Germany}

\begin{abstract}
    The interplay of spin-orbit coupling (SOC) and magnetism gives rise to a plethora of charge-to-spin conversion phenomena that harbor great potential for spintronics applications. In addition to the spin Hall effect, magnets may exhibit a magnetic spin Hall effect (MSHE), as was recently discovered [Kimata \textit{et al.}, Nature \textbf{565}, 627-630 (2019)]. To date, the MSHE is still awaiting its intuitive explanation.  Here we relate the MSHE to the vorticity of spin currents in the Fermi sea, which explains pictorially the origin of the MSHE\@. For all magnetic Laue groups that allow for nonzero spin current vorticities the related tensor elements of the MSH conductivity are given. Minimal requirements for the occurrence of a MSHE are compatibility with either a magnetization or a magnetic toroidal quadrupole. This finding implies in particular that the MSHE is expected in all ferromagnets with sufficiently large SOC\@. To substantiate our symmetry analysis, we present various models, in particular a two-dimensional magnetized Rashba electron gas, that corroborate an interpretation by means of spin current vortices. Considering thermally induced spin transport and the magnetic spin Nernst effect in magnetic insulators, which are brought about by magnons, our findings for electron transport can be carried over to the realm of spincaloritronics, heat-to-spin conversion, and energy harvesting.
\end{abstract}

\date{\today}

\maketitle

\section{From the conventional to the magnetic spin Hall effect}
The spin Hall effect (SHE) \cite{Sinova2015} and its inverse are without doubt important discoveries \cite{Kato2004, Wunderlich2005, Saitoh2006, Valenzuela2006, Zhao2006} in the field of spintronics \cite{wolf2001spintronics, Zutic2004}. They serve not only as `working horses' for generating and detecting spin currents \cite{jungwirth2012spin} but also as key ingredients in spin-orbit torque devices for electric magnetization switching \cite{chernyshov2009evidence, Liu2011, Liu2012}. Compared to spin-transfer torque devices \cite{Slonczewski1996, Berger1996, Ralph2008, Brataas2012, Khvalkovskiy2013}, spin-orbit torque devices are faster, more robust, and consume less power upon operation \cite{Gambardella2011, Jabeur2014, Prenat2016}; for a recent review see Ref.~\onlinecite{Li2018magmani}.

While the anomalous Hall effect (AHE) in a magnet \cite{Nagaosa2010} produces a transverse charge current density upon applying an electric field $\vec{E}$, the SHE in a nonmagnet produces a transverse spin current density $\langle \vec{j}^\gamma \rangle = \underline{\sigma}^\gamma \vec{E}$ ($\gamma = x, y, z$ indicates the transported spin component). Mathematically, the SHE is quantified by the antisymmetric part of the spin conductivity tensor $\underline{\sigma}^\gamma$. For example, the $\sigma^z_{xy}$ element comprises $z$-polarized spin currents in $x$ direction as a response to an electric field in $y$ direction.

In a simple picture, the intrinsic SHE \cite{Murakami2003, SinovaUniversal2004} is explained by spinning electrons that experience a spin-dependent Magnus force caused by spin-orbit coupling (SOC). It appears as if `built-in' spin-dependent magnetic fields evoke spin-dependent Lorentz forces that result in a transverse pure spin current. The extrinsic SHE \cite{Dyakonov1971, Hirsch1999, ZhangSHE2000} is covered by Mott scattering at defects \cite{Mott1929}.

Since the SHE does not rely on broken time-reversal symmetry (TRS), it is featured in nonmagnetic metals \cite{Guo2008} or semiconductors \cite{Kato2004}. Imposing few demands on a material's properties, a SHE can be expected in any material with sufficiently large SOC (or, instead of SOC, with a noncollinear magnetic texture \cite{zhang2018spin}). From a mathematical perspective the existence of an SHE can be traced to the transformation behavior of the sum
\begin{align}
     \sigma_{xy}^z-\sigma_{yx}^z + \sigma_{yz}^x - \sigma_{zy}^x + \sigma_{zx}^y - \sigma_{xz}^y, \label{eq:SHEtraditional}
\end{align}
of antisymmetric spin conductivity tensor elements traditionally associated with a SHE (applied field, current flow direction, and transported spin component are mutually orthogonal). Taking time-reversal evenness for granted, this sum behaves like an electric monopole (space-inversion even, scalar). For there are no crystalline symmetries (reflections, rotations, inversions) that could render such an object zero, a SHE can basically occur in any material. For the rest of this Paper, we refer to this SHE as `conventional SHE.'

To combine the virtues of transverse spin transport with magnetic recording, the conventional SHE was studied in magnetic materials with broken TRS (ferromagnets \cite{Miao2013, Taniguchi2015, Tian2016, Wu2017ferro, Das2017, Humphries2017, Amin2018, Gibbons2018, Bose2018, Baek2018, Omori2019, Amin2019, Qu2019ArXiv, Wang2019anomaloustorque} or antiferromagnets \cite{Zelezny2017, vzelezny2018spin, zhang2018spin, Chen2018, Kimata2019}), which revealed various phenomena associated with the interplay of SOC and magnetism. 
For example, ferromagnetic metals exhibit an (inverse) conventional SHE \cite{Miao2013}; unaffected by magnetization reversal \cite{Tian2016} it is time-reversal even. 

Since charge currents in ferromagnets are intrinsically spin-polarized, transverse AHE currents are spin-polarized as well and are used to generate spin torques \cite{Taniguchi2015, Das2017}. This effect is sometimes referred to as anomalous SHE \cite{Das2017}, but is fundamentally a conventional SHE\@.
Since these spin currents are tied to the AHE charge currents, the spin accumulations brought about by this effect can be manipulated by varying the magnetization direction \cite{Gibbons2018, Amin2019, Qu2019ArXiv}. This finding can be understood by considering symmetries.
For a nonmagnetic cubic material only the components in Eq.~\eqref{eq:SHEtraditional} are allowed nonzero. In contrast, a ferromagnetic material magnetized along a generic direction has a lower symmetry: there are no constraints that prohibit `populating' the entire spin conductivity tensor. Upon manipulation of the magnetization, an electric field in, say, $y$ direction causes an arbitrary spin polarization flowing in, e.\,g., $x$ direction. This offers greater versatility for spin torque applications than the conventional SHE in nonmagnetic cubic materials (nonmagnetic and noncubic materials also admit of greater versatility and nontraditional tensor elements \cite{Wimmer2015, Seemann2015, MacNeill2016, Zhou2019}, but they do not offer external means, such as magnetization, to manipulate spin polarizations).

Since TRS is intrinsically broken in magnets, one expects that spin accumulations brought about by transverse spin currents have two components, one that does not reverse under magnetization reversal (we will refer this effect as SHE, a subset of which is the conventional SHE) and a second that is reversed under magnetization reversal. This opposite behavior under time reversal causes different restrictions imposed by the magnetic point-group symmetry \cite{Seemann2015,Zelezny2017} on the two types of spin accumulations. In particular, the latter magnetism-induced accumulations do not have to be parallelly polarized to the SHE spin accumulations. Such signatures were observed in Ref.~\onlinecite{Humphries2017}.

The disentanglement of spin current contributions odd or even under time reversal has been elucidated in Ref.~\onlinecite{Zelezny2017}. In essence, the spin conductivity tensor in Kubo transport theory is decomposed into a time-reversal even part and a time-reversal odd part. Upon disregarding spin-dependent scattering, skew scattering, and side jumps, the time-reversal even part is associated with `intrinsic' contributions to the spin conductivity, a contribution given solely in terms of band structure properties (in the so-called clean limit) \cite{Zelezny2017}. Likewise, the time-reversal odd part is associated with `extrinsic' contributions that depend on relaxation times \cite{Zelezny2017}. The latter gives rise to the magnetism-induced effects. In systems with low symmetry both parts contribute to all components of $\underline{\sigma}^\gamma$ and, in particular, to its antisymmetric part: the time-reversal even part gives rise to the SHE and the odd part to the \emph{magnetic} spin Hall effect (MSHE) \cite{Zelezny2017,Chen2018, Kimata2019}.

The MSHE has recently been experimentally detected in the noncollinear antiferromagnet Mn$_3$Sn \cite{Kimata2019}, and the aforementioned results of Ref.~\onlinecite{Humphries2017} on ferromagnets can also be considered proof of the MSHE (in Ref.~\onlinecite{Davidson2019} referred to as `transverse SHE with spin rotation').
Although instances of the MSHE have been identified, an intuitive picture that explains how and under which circumstances this effect comes about is missing.

\section{Chiral vortices of spin currents: Summary of this Paper}
\label{sec:overview}
We offer a vivid microscopic picture of the MSHE by relating it to the spin current vorticity (SCV) of the Fermi sea or, equivalently, to the circulation of spin currents about the Fermi surface.

In a rough draft, magnetic materials feature spin current whirlpools (or vortices) in reciprocal space for each of the three spin directions $\gamma = x, y, z$; as usual for angular quantities, we denote the axis of a vortex by a vector $\vec{\omega}^\gamma$. Similar to water whirlpools (in real space), whose handedness leads to an asymmetric deflection of plane water waves, the spin current whirlpools (in reciprocal space) cause an asymmetric deflection of the respective spin component. Since the spin current vortices occur in reciprocal space, they are delocalized in real space and, hence, do not act as scattering centers (like defects) but rather like an overall vortical background. 
To rephrase this statement in mathematical terms: although spin transport is treated within the constant relaxation time approximation that does not capture asymmetric scattering at defects (thereby ruling out extrinsic skew scattering and side jump contributions), the MSHE is captured, because the spin current itself---but not the scattering---is chiral.

In terms of SCVs, the time-reversal odd nature of the MSHE is easily understood as a reversal of a vortex's handedness that results in opposite deflection. Then, a reversal of the magnetic texture has to reverse the spin accumulations brought about by the MSHE spin currents as well; recall that SHE spin currents remain unaffected.

In order to show that these spin current vortices may exist we analyze all magnetic Laue groups (MLGs) with respect to their compatibility with a nonzero SCV, thereby identifying all possible MSHE scenarios.

One especially simple scenario is a ferromagnet with SOC: assuming a tetragonal ferromagnet with magnetization $\vec{M}$ in $z$ direction we find the SCVs
\begin{align}
    \vec{\omega}^x \upuparrows \hat{\vec{y}}, 
    \quad
    \vec{\omega}^y \upuparrows -\hat{\vec{x}}, 
    \quad \text{and} \quad
    \vec{\omega}^z = \vec{0}. 
    \label{eq:vorticities-in-ferro}
\end{align}
For the $z$ spin component, the Fermi sea is loosely speaking `calm' and does not cause an MSHE\@. In contrast, the $x$ and $y$ spin components `experience a rough chiral Fermi sea': the nonzero vorticities cause MSHEs. More precisely, the MSHE  for the $x$ ($y$) spin component takes place in the $xz$ ($yz$) plane. Consequently, such ferromagnets exhibit nonzero antisymmetric parts of $\sigma^\gamma_{\gamma z}$ (and $\sigma^\gamma_{ z \gamma}$) as long as $\gamma = x, y$; the transported spin component, the electric field, and the flow direction of the spin current lie within a plane that contains the magnetization. This is why the MSHE spin currents are pure: the transverse AHE charge currents compatible with a magnetization in $z$ direction flow within the $xy$ plane (i.\,e., normal to the magnetization).

\begin{figure}
    \centering
    \includegraphics[width=1\columnwidth]{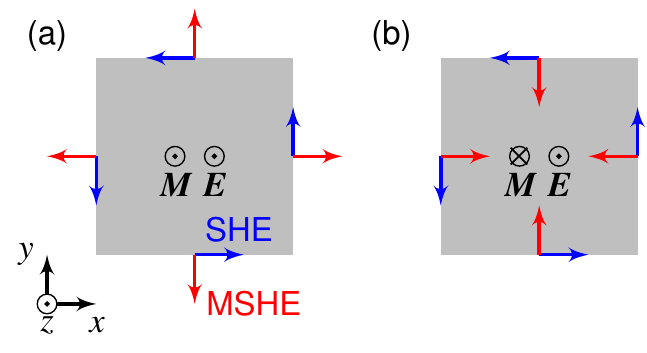}
    \caption{Conventional and magnetic spin Hall effect in ferromagnets with magnetization $\vec{M}$ along (a) $z$ direction and (b) $-z$ direction (e.\,g., MLG $4/mm'm'$). Upon application of an electric field $\vec{E}$ in $z$ direction (and a charge current in the same direction) spin accumulates at the boundaries of the sample. Those associated with the conventional spin Hall effect (SHE) are indicated by blue arrows, those with the MSHE by red arrows. Magnetization reversal acts only on the MSHE accumulations.}
    \label{fig:MSHEFerro}
\end{figure}

To elaborate on the difference to the SHE, let us assume that an electric field  $\vec{E} \parallel \vec{M} \parallel \hat{\vec{z}}$ is applied, as depicted in Fig.~\ref{fig:MSHEFerro}. Due to the (conventional) SHE spin conductivity tensor elements ($\sigma^x_{yz} = -\sigma^y_{xz}$) spin is accumulated within the surface planes of the sample (blue arrows). The polarization of these spin accumulations is orthogonal to both $\vec{E}$ and the surface normal. Being time-reversal even, it does not flip under magnetization reversal [compare panel (a) vs.\ (b)]. In contrast, the MSHE ($\sigma^x_{xz} = \sigma^y_{yz}$) causes additional accumulations polarized normal to the surface planes (red arrows). Their time-reversal odd nature forces a flip upon magnetization reversal, as is represented by the reversed red arrows in panel (b).

Thus, the MSHE allows to generate a spin accumulation orthogonal to the conventional SHE spin accumulation. In scenarios in which the magnetization of the ferromagnet is fixed this feature may result in the decisive spin accumulation direction necessary to perform a particular spin torque switching. For example, the field-free magnetization switching of a perpendicularly magnetized film observed in Ref.~\onlinecite{Baek2018} may be explained in terms of MSHE spin accumulations. We emphasize that the existence of spin current vortices is a bulk property that gives rise to bulk MSHE spin currents, which, in turn, cause spin accumulations at the edges or interfaces. Additional interface effects, as those accounted for in Refs.~\onlinecite{Amin2018}, \onlinecite{Baek2018}, and~\onlinecite{Hoenemann2019}, come on top.

That the SHE and MSHE cause spin accumulations pointing in orthogonal directions is a speciality of the MLGs $4/mm'm'$, $4'/mm'm$, and $m'm'm$, which allow for the `clearest' disentanglement of MSHE and SHE\@. For other MLGs there is at least one element of the spin conductivity tensor that carries simultaneously contributions from the SHE and the MSHE, leaving the behavior under time-reversal (texture reversal) as the only distinguishing characteristic.

Apart from three-dimensional ferromagnets, two-dimensional electron gases (2DEGs) appear highly attractive. 2DEGs are well known for their efficient charge-to-spin conversion due to large Rashba SOC \cite{Lesne2016}, magnetism in combination with superconductivity \cite{Brinkman2007, Reyren2007, Bert2011, Dikin2011, Li2011TwoDeg, joshua2012universal}, and electrical controllability \cite{Caviglia2008}. Recent progress in achieving room temperature magnetism in 2DEGs \cite{Zhang2018TwoDeg} suggests to investigate the MSHE in these systems. To do so we consider a minimal Rashba Hamiltonian with warping and an exchange field whose direction provides a handle to switch between different MLGs. It turns out that upon in-plane rotation of the field, the MSHE of in-plane polarized spins is manipulated but also that of out-of-plane polarized spins. Similar conclusions hold for topological Dirac surface states, as in Sn-doped Bi$_2$Te$_3$ \cite{Fu2009} (e.\,g., in the presence of exchange fields due to proximity to a ferromagnetic normal insulator \cite{Eremeev2013}), and for the noncollinear antiferromagnet Mn$_3$Sn.

Instead of an electric field, a temperature gradient may be utilized to cause thermodynamic non-equilibrium and spin transport. As above for the SHE and the MSHE, time-reversal even transverse spin transport is then referred to as spin Nernst effect (SNE), and the time-reversal odd partner is termed magnetic SNE (MSNE). Since the spin current vortices in reciprocal space exist irrespective of the driving force, the existence of an MSHE immediately implies that of an MSNE\@. Recalling that the symmetry analysis is independent of the type of spin carriers, it applies just as well to magnetic insulators, in which spin is transported by magnons. We present a proof of principle by considering antiferromagnetic spin textures, as in Mn$_3$Sn, and demonstrate that the magnetic excitations give rise to nonzero spin current vortices and, thus, to a magnonic MSNE\@. Therefore, the results of this Paper can be carried over to the realm of spincaloritronics in magnetic insulators, where they may inspire studies of novel heat-to-spin conversion mechanisms and energy harvesting concepts.

The remainder of the Paper is organized as follows. In Sec.~\ref{sec:MSHEgeneral} the theoretical framework within which we describe spin transport is introduced. We disentangle time-reversal even from odd contributions in Sec.~\ref{sec:spin_decomposition}, isolate the MSHE and introduce the SCV interpretation in Sec.~\ref{sec:MSHEisolating}. Then, we turn to the symmetry analysis of all MLGs and summarize key findings in Sec.~\ref{sec:symmetry_analysis}. These are elaborated on in Sec.~\ref{sec:examples} by considering specific toy models; the latter serve to underline the minimal requirements for a nonzero MSHE (Sec.~\ref{sec:minimal-req}), to make connection to magnetized Rashba materials (Sec.~\ref{sec:MSHEinRashba}, and to demonstrate the magnonic MSNE (Sec.~\ref{sec:MSNE}). 
We discuss the relation of our work to literature in Sec.~\ref{sec:Discussion} and summarize in Sec.~\ref{sec:conclusion}.

\section{Linear-response theory of the magnetic spin Hall effect}
\label{sec:MSHEgeneral}
The elements of the optical spin conductivity tensor read \cite{Seemann2015}
\begin{align}
    \sigma^\gamma_{\mu\nu}(\varpi) = \frac{1}{V} \int_0^\infty \mathrm{d} t\, \mathrm{e}^{\mathrm{i}\varpi t} \int_0^\beta \mathrm{d}\kappa \left\langle J_\nu J^\gamma_\mu(t+\mathrm{i} \hbar \kappa) \right\rangle \label{eq:fullKubo}
\end{align}
in Kubo linear-response theory \cite{Kubo1957,Mahan2000}. $J_\nu$, $J^\gamma_\mu$, $\varpi$, $V$, and $\beta$ are the total charge and spin current operators, the frequency of $\vec{E}(\varpi)$, the system's volume, and the inverse temperature, respectively. The shape of $\underline{\sigma}^\gamma$ was derived for all MLGs by symmetry arguments in Ref.~\onlinecite{Seemann2015}. However, such a superordinate symmetry approach neither provides insights into the character of the MSHE nor does it identify the MSHE contributions to the tensor elements. The latter requires to decompose $\underline{\sigma}^\gamma$. 

\subsection{Decomposition of the spin conductivity tensor}
\label{sec:spin_decomposition}
We work in the limit of non-interacting electrons described by the Hamiltonian 
\begin{align}
    H = \sum_{\vec{k}} \adj{\vec{\varPsi}}_{\vec{k}} \underline{H}_{\vec{k}} \vec{\varPsi}_{\vec{k}}
\end{align}
in crystal momentum ($\vec{k}$) representation. $\adj{\vec{\varPsi}}_{\vec{k}}$ ($\vec{\varPsi}_{\vec{k}}$) is a vector of electronic creation (annihilation) operators; its index runs over spin, orbitals, and basis lattice sites. The eigenenergies $\varepsilon_{n \vec{k}}$ and corresponding eigenvectors $|n\rangle = |u_{n \vec{k}} \rangle$, which represent the lattice-periodic part of a Bloch wavefunction with band index $n$, are obtained from diagonalizing the Hamilton kernel $\underline{H}_{\vec{k}}$.

The dc spin conductivity then reads
\begin{align}
      \sigma^\gamma_{\mu\nu} \equiv \mathrm{Re} \left[ \lim_{\varpi \to 0} \sigma^\gamma_{\mu\nu}(\varpi) \right] 
    = \sigma^{\gamma,\mathrm{odd}}_{\mu\nu} + \sigma^{\gamma,\mathrm{even}}_{\mu\nu},
\end{align}
with the two contributions \cite{Zelezny2017}
\begin{subequations}
\begin{align}
    \sigma^{\gamma,\mathrm{odd}}_{\mu\nu} &= \frac{\hbar^2 \varGamma}{V} \sum_{n,m,\vec{k}}
        \frac{f_{m \vec{k}} - f_{n \vec{k}}}{ \varepsilon_{n \vec{k}} - \varepsilon_{m \vec{k}} }
        \frac{ \mathrm{Re} \left( \langle n | \underline{J}^\gamma_{\vec{k},\mu} | m \rangle \langle m | \underline{J}_{\vec{k},\nu} | n \rangle \right) }{\left( \varepsilon_{n \vec{k}} - \varepsilon_{m \vec{k}} \right)^2 + \left( \hbar \varGamma\right)^2}, \label{eq:spin_cond_odd}
    \\
    \sigma^{\gamma,\mathrm{even}}_{\mu\nu} &= - \frac{\hbar}{V} \sum_{n,m,\vec{k}}
        \left( f_{m \vec{k}} - f_{n \vec{k}}\right)
        \frac{ \mathrm{Im} \left( \langle n | \underline{J}^\gamma_{\vec{k},\mu} | m \rangle \langle m | \underline{J}_{\vec{k},\nu} | n \rangle \right) }{\left( \varepsilon_{n \vec{k}} - \varepsilon_{m \vec{k}} \right)^2 + \left( \hbar \varGamma\right)^2}. \label{eq:spin_cond_even}
\end{align}
\end{subequations}
$f_{n \vec{k}} = (\mathrm{e}^{\beta( \varepsilon_{n \vec{k}} - \varepsilon_\mathrm{F})}+1)^{-1}$ is the Fermi distribution function with Fermi energy $\varepsilon_\mathrm{F}$.   $\hbar \varGamma$ is an artificial spectral broadening and the total currents are decomposed into their Fourier kernels $\underline{J}^\gamma_{\vec{k},\mu}$ and $\underline{J}_{\vec{k},\nu} = -e \underline{v}_{\vec{k},\nu} = -e \hbar^{-1} \partial \underline{H}_{\vec{k}}/\partial k_\nu$ (in the $\vec{\varPsi}_{\vec{k}}$ basis), respectively.

The superscripts of $\sigma^{\gamma,\mathrm{odd}}_{\mu\nu}$ and $\sigma^{\gamma,\mathrm{even}}_{\mu\nu}$ indicate their behavior under time-reversal. We recall that spin (charge) current is time-reversal even (odd) and that the TR operator comprises complex conjugation \cite{Zelezny2017}. The behavior under time reversal can be addressed by a reversal of the magnetic texture (a collection $\{\vec{m}_i\}$ of magnetic moments):
\begin{equation}
\begin{aligned}
    \sigma^{\gamma,\mathrm{odd}}_{\mu\nu} [\{\vec{m}_i\}] &= 
    -\sigma^{\gamma,\mathrm{odd}}_{\mu\nu} [-\{\vec{m}_i\}],
    \\
    \sigma^{\gamma,\mathrm{even}}_{\mu\nu} [\{\vec{m}_i\}] &= 
    \sigma^{\gamma,\mathrm{even}}_{\mu\nu} [-\{\vec{m}_i\}].
\end{aligned}
\end{equation}
In an experiment these contributions can be disentangled by measuring the spin accumulations brought about by the spin currents for both the original and the reversed texture.

Following up on Eq.~\eqref{eq:spin_cond_odd}, a MSHE was identified by symmetry arguments for Mn$_3 X$ ($X=\mathrm{Sn,Ga,Ge}$) in Ref.~\onlinecite{Zelezny2017} (see the first entry in the right column of Tab.~I of that paper and consider $\sigma^x_{xy} \ne \sigma^x_{yx}$, which makes the antisymmetric part of $\underline{\sigma}^x$ nonzero). The term MSHE was coined in Refs.~\onlinecite{Chen2018, Kimata2019}, the latter of which reported on its experimental observation in Mn$_3$Sn. 

In what follows we concentrate on $\sigma^{\gamma,\mathrm{odd}}_{\mu\nu}$, because $\sigma^{\gamma,\mathrm{even}}_{\mu\nu}$ is related to the intrinsic SHE \cite{SinovaUniversal2004} which is of minor interest in this Paper. We decompose $\sigma^{\gamma,\mathrm{odd}}_{\mu\nu}$ into intraband contributions ($n = m$)
\begin{align}
    \sigma^{\gamma,\mathrm{odd}}_{\mu\nu,\mathrm{intra}} &= \frac{1}{\varGamma V} \sum_{n, \vec{k}}
        J^\gamma_{n \vec{k},\mu} J_{ n \vec{k},\nu} 
        \left( - \frac{\partial f_{n \vec{k}}}{\partial \varepsilon} \right) \label{eq:spin_cond_odd_intra}
\end{align}
and interband contributions given by Eq.~\eqref{eq:spin_cond_odd} with the sum restricted to $n \ne m$. $J^\gamma_{ n \vec{k},\mu} \equiv \langle n | \underline{J}^\gamma_{\vec{k},\mu} | n \rangle $ is the spin and $J_{n \vec{k},\nu} \equiv - e \langle n | \underline{v}_{\vec{k},\nu} | n \rangle = -e v_{n \vec{k},\nu}$ the charge current expectation value; $v_{n \vec{k},\nu} = \hbar^{-1} \partial \varepsilon_{n \vec{k}} / \partial k_\nu$ is the group velocity and $e > 0$ the elementary charge. We note in passing that Eq.~\eqref{eq:spin_cond_odd_intra} can also be derived within the semiclassical Boltzmann transport theory, assuming a constant relaxation time.
For $\varGamma \to 0$, $\sigma^{\gamma,\mathrm{odd}}_{\mu\nu,\mathrm{intra}}$ diverges while the interband contributions converge to a constant. From here on, we thus drop the specifiers `odd' and `intra', work in the `almost clean' limit ($\hbar \varGamma > 0$ is the smallest energy scale), and focus on the dominating intraband contribution in Eq.~\eqref{eq:spin_cond_odd_intra}. 

\subsection{Identification of the MSHE contributions}
\label{sec:MSHEisolating}
The antisymmetric part
\begin{align}
    \underline{\sigma}^{\gamma,\mathrm{(a)}} \equiv \frac{\underline{\sigma}^{\gamma} - \underline{\sigma}^{\gamma, \mathrm{T}}}{2}
\end{align}   
of the time-odd spin conductivity tensor describes the MSHE\@. Since any antisymmetric matrix can be represented by a vector, we introduce the $\gamma$-spin MSHE vector as $[\vec{\sigma}^{\gamma}_\mathrm{MSHE}]_\times = \underline{\sigma}^{\gamma,\mathrm{(a)}}$, written compactly as [cf.~eq.~\eqref{eq:spin_cond_odd_intra}]
\begin{align}
    \vec{\sigma}^{\gamma}_\mathrm{MSHE} 
    \equiv
    \begin{pmatrix}
    \sigma^{\gamma,\mathrm{(a)}}_{yz} \\
    \sigma^{\gamma,\mathrm{(a)}}_{zx} \\
    \sigma^{\gamma,\mathrm{(a)}}_{xy}
    \end{pmatrix}
    =
    \frac{1}{2\varGamma V} \sum_{n, \vec{k}} \vec{J}^\gamma_{n \vec{k}} \times \vec{J}_{n \vec{k}}  \left( - \frac{\partial f_{n \vec{k}}}{\partial \varepsilon} \right).
    \label{eq:MSHEvector}
\end{align}
At zero temperature, $ - \partial f_{n \vec{k}} / \partial \varepsilon = \delta( \varepsilon_{n \vec{k}} - \varepsilon_\mathrm{F})$ allows to replace the $\vec{k}$ summation by an integral over the Fermi surface,
\begin{align}
     \vec{\sigma}^\gamma_\mathrm{MSHE} = \frac{e}{2 \hbar \varGamma (2\pi)^3 } \sum_{n} \oint_{\varepsilon_n = \varepsilon_\mathrm{F}} \hat{\vec{v}}_{n\vec{k}} \times \vec{J}^\gamma_{n \vec{k}}\, \mathrm{d} S \label{eq:MSHE_cond}
\end{align}
($\hat{\vec{v}}_{n\vec{k}} = \vec{v}_{n \vec{k}} / v_{n\vec{k}}$ is the local normal of the Fermi surface).
This integral measures the tangential vector flow of $\vec{J}^\gamma_{n \vec{k}}$ on the Fermi surface. $\vec{\sigma}^\gamma_\mathrm{MSHE}$ is nonzero if there is an integrated sense of rotation of the spin current about the Fermi surface. 

Alternatively, we write Eq.~\eqref{eq:MSHE_cond} as a Fermi sea integral,
\begin{align}
     \vec{\sigma}^\gamma_\mathrm{MSHE} = \frac{e}{2 \hbar \varGamma (2\pi)^3 } \vec{\omega}^\gamma (\varepsilon_\mathrm{F}), \label{eq:MSHE_cond_vorticity}
\end{align}
over the net spin current vorticity (SCV)
\begin{align}
    \vec{\omega}^\gamma (\varepsilon_\mathrm{F}) \equiv \sum_{n} \iiint_{\varepsilon_n \le \varepsilon_\mathrm{F}} \vec{\omega}^\gamma_{n \vec{k}} \, \mathrm{d}^3 k,
    \quad
    \vec{\omega}^\gamma_{n \vec{k}} \equiv \vec{\nabla}_{\vec{k}} \times  \vec{J}^\gamma_{n \vec{k}}, \label{eq:vorticity-definition}
\end{align}
that is defined in analogy to the vorticity of a fluid \cite{LandauLifshitz6}. $\vec{\omega}^\gamma_{n \vec{k}}$ describes the local rotation, shear or curvature of $\vec{J}^\gamma_{n \vec{k}}$. Figuratively speaking, the vorticity of a vector field is nonzero at those points at which a paddle wheel would start to rotate (note that integrals over fully occupied bands are zero, i.\,e., each band has a vanishing total SCV).

Equations~\eqref{eq:MSHE_cond} -- \eqref{eq:vorticity-definition} are our main findings. They show that a MSHE is a result of the spin current circulation about the Fermi surface [Eq.~\eqref{eq:MSHE_cond}] or, put differently, a result of a finite SCV in the Fermi sea [Eq.~\eqref{eq:MSHE_cond_vorticity}]. 

For illustration we stretch the analogy to fluid vortices and recall the time-reversal asymmetric propagation of an acoustic wave through a fluid with a vortex \cite{roux1995experimental}, briefly laid out in the introduction. The broken TRS in magnets causes SCVs $\vec{\omega}^\gamma$ for each spin component $\gamma = x, y, z$ in the Fermi sea, which is experienced by the $\gamma$ spin component of an electron's Bloch wave propagating through the crystal. A consequence is a Hall-like deflection within the plane normal to $\vec{\omega}^\gamma$ of that spin component. Time reversal is equivalent to inversion of the vortex's circulation direction ($\vec{\omega}^\gamma \to - \vec{\omega}^\gamma$), which signifies the time-odd signature of the MSHE\@.

\begin{figure}
    \centering
    \includegraphics[width=1\columnwidth]{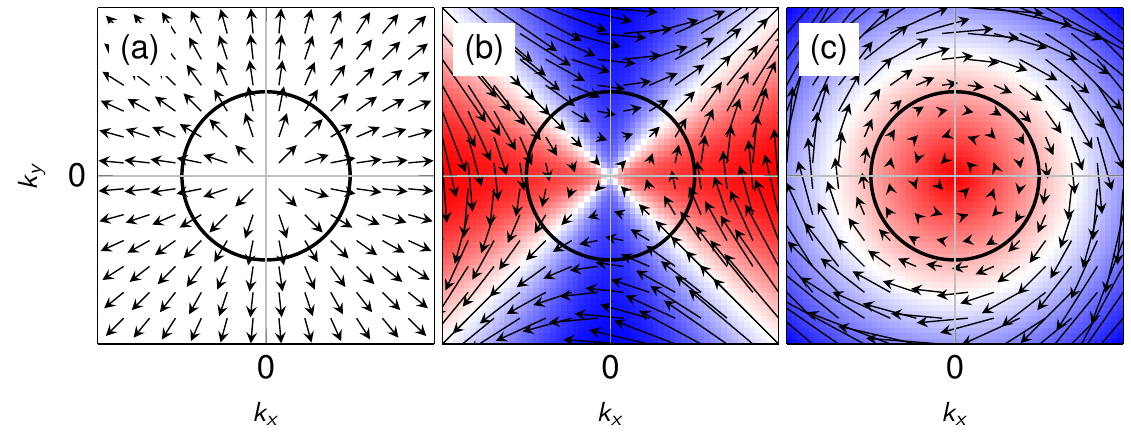}
    \caption{Vorticities of spin current vector fields in reciprocal space. Red/white/blue color indicates positive/zero/blue vorticity in a region about the origin. The integral over the vorticity within the Fermi surface, indicated by black circles, is proportional to the magnetic spin Hall conductivity. (a) The irrotational source field has zero vorticity. (b) The quadrupolar field has locally nonzero vorticity but zero integral. (c) A general vortex with vorticity of varying sign.}
    \label{fig:MSHEKSpace}
\end{figure}

Considering reciprocal space, a simple picture may be helpful. In a two-dimensional crystal with a single Fermi line, the spin current vector field $\vec{J}^\gamma_{\vec{k}}$ may look as depicted in Fig.~\ref{fig:MSHEKSpace} (we suppressed the band index). The integral of the $\vec{k}$-dependent vorticity over the Fermi sea is proportional to the magnetic spin Hall conductivity. In scenario (a), $ \vec{J}^\gamma_{\vec{k}}$ is irrotational and, thus, has zero vorticity. In (b) $\vec{J}^\gamma_{\vec{k}}$ shows local vorticity that integrates to zero due to symmetry. And in (c), the Fermi surface cuts out a region with nonzero vorticity causing a MSHE\@. To check the behavior under time reversal recall the mapping $\vec{J}^\gamma_{\vec{k}}$ to $\vec{J}^\gamma_{-\vec{k}}$, which reverses the circulation direction. 

One may discuss the effect in terms of a shift of the Fermi surface that is caused by the redistribution of electrons. An electric field $\vec{E}$ along $-x$ direction produces a shift in positive $k_x$ direction (accounting for the negative electron charge). For the situation in Fig.~\ref{fig:MSHEKSpace}(a), this displacement does not yield a transversal response since $\langle \vec{j}^\gamma \rangle \parallel \hat{\vec{x}}$. However, for (b) and (c) $\langle \vec{j}^\gamma \rangle \parallel \hat{\vec{y}}$ and $\langle \vec{j}^\gamma \rangle \parallel -\hat{\vec{y}}$, respectively. If $\vec{E}$ is along the $-y$ direction (shift in positive $k_y$ direction),  $\langle \vec{j}^\gamma \rangle \parallel \hat{\vec{x}}$ for both (b) and (c). Hence, only a finite SCV, as depicted in (c), fixes the sign of $\langle \vec{j}^\gamma \rangle \times \vec{E}$, thereby causing a nonzero antisymmetric part of $\underline{\sigma}^\gamma$, that is a MSHE\@. The scenario (b) gives rise to a symmetric part of $\underline{\sigma}^\gamma$, conceivably referred to as `planar magnetic spin Hall effect'.

\subsection{Symmetry analysis}
\label{sec:symmetry_analysis}
Although breaking of TRS is necessary for a nonzero local SCV $\vec{\omega}^\gamma_{n\vec{k}}$, it is not sufficient because symmetries of the magnetic crystal may render the SCV integral in Eq.~\eqref{eq:vorticity-definition} zero [cf.\ Fig.~\ref{fig:MSHEKSpace}(b)]. In what follows we derive which MLGs do or do not allow for a MSHE\@. The restriction to MLGs---instead to magnetic point groups---is feasible because $\underline{\sigma}^\gamma$ is related to the correlation function of a spin current and a charge current [Eq.~\eqref{eq:fullKubo}]. Both currents change sign upon inversion; thus, the presence or absence of inversion symmetry does not impose restrictions on the shape of the spin conductivity tensor. One may then augment each magnetic point group with the element of space inversion to map it onto the set of MLGs. Recall that the considerably smaller number of MLGs facilitates the analysis. This argumentation is in line with Refs.~\cite{Kleiner1966, Kleiner1967, Kleiner1969, Seemann2015}. We like to refer the reader to Ref.~\cite{Seemann2015} for mappings of magnetic point groups onto MLGs.

We combine the three spin-dependent MSHE vectors of Eq.~\eqref{eq:MSHE_cond_vorticity} to the MSHE tensor
\begin{align}
    \underline{\sigma}_\mathrm{MSHE}  \equiv \left( \vec{\sigma}^x_\mathrm{MSHE}, \vec{\sigma}^y_\mathrm{MSHE}, \vec{\sigma}^z_\mathrm{MSHE} \right) 
    \propto
    \underline{\omega}  \equiv \left( \vec{\omega}^x, \vec{\omega}^y, \vec{\omega}^z \right). \label{eq:connectionMSHEvorticity}
\end{align}
Equation~\eqref{eq:connectionMSHEvorticity} links the elements of $\underline{\sigma}_\mathrm{MSHE}$ to those of the SCV tensor $\underline{\omega}$, the latter itself constructed from the three spin current vorticity vectors given in Eq.~\eqref{eq:vorticity-definition} (argument  $\varepsilon_\mathrm{F}$ suppressed). 
$\underline{\omega}$ can be decomposed into three contributions:
\begin{align}
    \underline{\omega} = \omega \underline{I} + [\vec{\varOmega}]_\times + \underline{W}, \label{eq:vorticity-decomp}
\end{align}
that is a scalar $\omega = \operatorname{Tr}(\underline{\omega}) / 3$, a vector $[\vec{\varOmega}]_\times = ( \underline{\omega} - \underline{\omega}^\mathrm{T} )/2$, and a traceless symmetric tensor $\underline{W} = ( \underline{\omega} + \underline{\omega}^\mathrm{T} )/2 - \omega \underline{I}$, with $\underline{I}$ the $3 \times 3$ identity matrix. Recalling that $\underline{\omega}$ is time-reversal odd but space-inversion even and calling to mind the transformation properties of electromagnetic multipoles \cite{Nanz2016}, one finds that $\omega$, $\vec{\varOmega}$, and $\underline{W}$ transform as a magnetic toroidal monopole (Jahn symbol $a$, Ref.~\onlinecite{Jahn1949}), a magnetic dipole ($eaV$), and a magnetic toroidal quadrupole ($a[V^2]$), respectively. Such a multipole decomposition is in line with Ref.~\onlinecite{Hayami2018} [cf.\ Eqs.~(D20) and (D21) of that publication].

Utilizing the \textsc{Mtensor} application \cite{Gallego2019} of the \textit{Bilbao Crystallographic Server} \cite{Aroyo2006, Aroyo2006b, aroyo2011crystallography}, we identified all MLGs permitting these multipoles. By virtue of Eqs.~\eqref{eq:MSHEvector} and~\eqref{eq:connectionMSHEvorticity} these results are carried forward to $\underline{\sigma}_\mathrm{MSHE}$  and $\underline{\sigma}^{\gamma,\mathrm{(a)}}$; a summary is given in Table~\ref{tab:table}. The results of the symmetry analysis are not restricted to the intraband approximation (i.\,e., the interpretation in terms of SCVs) but apply to Eq.~\eqref{eq:spin_cond_odd} as well [one may consider the summand in Eq.~\eqref{eq:spin_cond_odd} as a generalization of the SCV to interband contributions]. We now list and discuss illustrative key findings. 
  
\begin{table*}
    \centering
    \caption{Symmetry analysis of the MSHE\@. The columns give magnetic Laue groups (MLGs), components of the spin current vorticity (SCV) $\underline{\omega}$, and elements of spin conductivity $\underline{\sigma}^\gamma$ that are compatible with an MSHE\@.}
    \begin{ruledtabular}
    \begin{tabular}{c c c c c c c}
        \multirow{2}{*}{MLG\footnote{The MLGs $\bar{1}1'$, $2/m1'$, $mmm1'$, $4/m1'$, $6/m1'$, $4/mmm1'$, $6/mmm1'$, $\bar{3}1'$, $\bar{3}m1'$, $m\bar{3}1'$, and $m\bar{3}m1'$ contain pure time-reversal symmetry and are incompatible with a spin current vorticity and, thus, a MSHE\@. The MLGs $6'/m'$, $6'/m'mm'$, and $m\bar{3}m'$ are MSHE-incompatible as well.}} & \multicolumn{3}{c}{Admitted elements of SCV tensor $\underline{\omega}$}  & \multicolumn{3}{c}{MSHE part of spin conductivity tensor $\underline{\sigma}^\gamma$}  \\
        \cmidrule{2-4} \cmidrule{5-7}
         & $\omega$ & $\vec{\varOmega}$ & $\underline{W}$\footnote{Since $\underline{W}$ is symmetric, admittance of $W_i^j$ implies admittance of $W_j^i$.} & $\underline{\sigma}^{x,\mathrm{(a)}}$ & $\underline{\sigma}^{y,\mathrm{(a)}}$ & $\underline{\sigma}^{z,\mathrm{(a)}}$ \\
        \cmidrule{1-7}
        $\bar{1}$   & \multirow{1}{*}{$\omega$}  & $\varOmega_x$, $\varOmega_y$, $\varOmega_z$ & $W_x^x$, $W_y^y$, $W_z^z$, $W_x^y$, $W_x^z$,  $W_y^z$
        & $\begin{pmatrix} 0 & \sigma^x_{xy} & \sigma^x_{xz} \\  -\sigma^x_{xy} & 0 & \sigma_{yz}^x \\ -\sigma^x_{xz} & -\sigma_{yz}^x & 0 \end{pmatrix}$
        & $\begin{pmatrix} 0 & \sigma^y_{xy} & \sigma_{xz}^y \\  -\sigma^y_{xy} & 0 & \sigma^y_{yz} \\ -\sigma_{xz}^y & -\sigma^y_{yz} & 0 \end{pmatrix}$  
        & $\begin{pmatrix} 0 & \sigma_{xy}^z & \sigma^z_{xz} \\  -\sigma_{xy}^z & 0 & \sigma^z_{yz} \\ -\sigma^z_{xz} & -\sigma^z_{yz} & 0 \end{pmatrix}$
        \\
        \cmidrule{1-7}
        $2/m$   & \multirow{1}{*}{$\omega$} & $\varOmega_y$ &  $W_x^x$, $W_y^y$, $W_z^z$, $W_x^z$ 
        & $\begin{pmatrix} 0 & \sigma^x_{xy} & 0 \\  -\sigma^x_{xy} & 0 & \sigma_{yz}^x \\ 0 & -\sigma_{yz}^x & 0 \end{pmatrix}$
        & $\begin{pmatrix} 0 & 0 & \sigma_{xz}^y \\  0 & 0 & 0 \\ -\sigma_{xz}^y & 0 & 0 \end{pmatrix}$  
        & $\begin{pmatrix} 0 & \sigma_{xy}^z & 0 \\  -\sigma_{xy}^z & 0 & \sigma^z_{yz} \\ 0 & -\sigma^z_{yz} & 0 \end{pmatrix}$
        \\
        \cmidrule{1-7}
        $mmm$   & $\omega$ & $-$ &  $W_x^x$, $W_y^y$, $W_z^z$ 
        & $\begin{pmatrix} 0 & 0 & 0 \\  0 & 0 & \sigma_{yz}^x \\ 0 & -\sigma_{yz}^x & 0 \end{pmatrix}$
        & $\begin{pmatrix} 0 & 0 & \sigma_{xz}^y \\  0 & 0 & 0 \\ -\sigma_{xz}^y & 0 & 0 \end{pmatrix}$  
        & $\begin{pmatrix} 0 & \sigma_{xy}^z & 0 \\  -\sigma_{xy}^z & 0 & 0 \\ 0 & 0 & 0 \end{pmatrix}$
        \\
        \cmidrule{1-7}
        $4/m$   & \multirow{3}{*}{$\omega$}& \multirow{3}{*}{$\varOmega_z$} & \multirow{3}{*}{ $W_x^x=W_y^y$, $W_z^z$} 
        & \multirow{3}{*}{$\begin{pmatrix} 0 & 0 & \sigma_{xz}^x \\  0 & 0 & \sigma_{yz}^x \\ -\sigma_{xz}^x & -\sigma_{yz}^x & 0 \end{pmatrix}$}
        & \multirow{3}{*}{$\begin{pmatrix} 0 & 0 & -\sigma_{yz}^x \\  0 & 0 & \sigma_{xz}^x \\ \sigma_{yz}^x & -\sigma_{xz}^x & 0 \end{pmatrix}$ } 
        & \multirow{3}{*}{$\begin{pmatrix} 0 & \sigma_{xy}^z & 0 \\  -\sigma_{xy}^z & 0 & 0 \\ 0 & 0 & 0 \end{pmatrix}$}\\
        $6/m$   &  & &  \\
        $\bar{3}$   &  &  &  
        \\
        \cmidrule{1-7}
        $4/mmm$   & \multirow{3}{*}{$\omega$} & \multirow{3}{*}{$-$} & \multirow{3}{*}{$W_x^x=W_y^y$, $W_z^z$} 
        & \multirow{3}{*}{$\begin{pmatrix} 0 & 0 & 0 \\  0 & 0 & \sigma_{yz}^x \\ 0 & -\sigma_{yz}^x & 0 \end{pmatrix}$}
        & \multirow{3}{*}{$\begin{pmatrix} 0 & 0 & -\sigma_{yz}^x \\  0 & 0 & 0 \\ \sigma_{yz}^x & 0 & 0 \end{pmatrix}$ } 
        & \multirow{3}{*}{$\begin{pmatrix} 0 & \sigma_{xy}^z & 0 \\  -\sigma_{xy}^z & 0 & 0 \\ 0 & 0 & 0 \end{pmatrix}$}
        \\
        $6/mmm$   &  &  & \\
        $\bar{3}m$   &  &  & \\
        \cmidrule{1-7}
        $m\bar{3}$   & \multirow{3}{*}{$\omega$} &  \multirow{3}{*}{$-$}  &  \multirow{3}{*}{$W_x^x=W_y^y=W_z^z$} & 
        \multirow{3}{*}{$\begin{pmatrix} 0 & 0 & 0 \\  0 & 0 & \sigma_{yz}^x \\ 0 & -\sigma_{yz}^x & 0 \end{pmatrix}$}
        & \multirow{3}{*}{$\begin{pmatrix} 0 & 0 & -\sigma_{yz}^x \\  0 & 0 & 0 \\ \sigma_{yz}^x & 0 & 0 \end{pmatrix}$ } 
        & \multirow{3}{*}{$\begin{pmatrix} 0 & \sigma_{yz}^x & 0 \\  -\sigma_{yz}^x & 0 & 0 \\ 0 & 0 & 0 \end{pmatrix}$}  \\ 
        & & & & & & \\
        $m\bar{3}m$  &  &  &    \\
        \cmidrule{1-7}
        $2'/m'$  & \multirow{1}{*}{$-$} &  $\varOmega_x$, $\varOmega_z$ & $W_x^y$, $W_y^z$ 
        & $\begin{pmatrix} 0 & 0 & \sigma_{xz}^x \\  0 & 0 & 0 \\ -\sigma_{xz}^x & 0 & 0 \end{pmatrix}$
        & $\begin{pmatrix} 0 & \sigma_{xy}^y & 0 \\  -\sigma_{xy}^y & 0 & \sigma_{yz}^y \\ 0 & -\sigma_{yz}^y & 0 \end{pmatrix}$  
        & $\begin{pmatrix} 0 & 0 & \sigma_{xz}^z \\  0 & 0 & 0 \\ -\sigma_{xz}^z & 0 & 0 \end{pmatrix}$ 
        \\
        \cmidrule{1-7}
        $m'm'm$  & \multirow{1}{*}{$-$}&  $\varOmega_z$ & $W_x^y$ 
        & $\begin{pmatrix} 0 & 0 & \sigma_{xz}^x \\  0 & 0 & 0 \\ -\sigma_{xz}^x & 0 & 0 \end{pmatrix}$
        & $\begin{pmatrix} 0 & 0 & 0 \\  0 & 0 & \sigma_{yz}^y \\ 0 & -\sigma_{yz}^y & 0 \end{pmatrix}$  
        & $-$
        \\
        \cmidrule{1-7}
        $4'/m$  & \multirow{1}{*}{$-$} & $-$ & $W_x^y$, $W_x^x = - W_y^y$ 
        & $\begin{pmatrix} 0 & 0 & \sigma_{xz}^x \\  0 & 0 & \sigma_{yz}^x \\ -\sigma_{xz}^x & -\sigma_{yz}^x & 0 \end{pmatrix}$
        & $\begin{pmatrix} 0 & 0 & \sigma_{yz}^x \\  0 & 0 & -\sigma_{xz}^x \\ -\sigma_{yz}^x & \sigma_{xz}^x & 0 \end{pmatrix}$  
        & $-$
        \\
        \cmidrule{1-7}
        $4'/mm'm$  & \multirow{1}{*}{$-$} &  $-$ & $W_x^y$ 
        & $\begin{pmatrix} 0 & 0 & \sigma_{xz}^x \\  0 & 0 & 0 \\ -\sigma_{xz}^x & 0 & 0 \end{pmatrix}$
        & $\begin{pmatrix} 0 & 0 & 0 \\  0 & 0 & -\sigma_{xz}^x \\ 0 & \sigma_{xz}^x & 0 \end{pmatrix}$  
        & $-$
        \\
        \cmidrule{1-7}
        $4/mm'm'$  & \multirow{3}{*}{$-$}  & \multirow{3}{*}{$\varOmega_z$} & \multirow{3}{*}{$-$} & \multirow{3}{*}{$\begin{pmatrix} 0 & 0 & \sigma_{xz}^x \\  0 & 0 & 0 \\ -\sigma_{xz}^x & 0 & 0 \end{pmatrix}$} 
        & \multirow{3}{*}{$\begin{pmatrix} 0 & 0 & 0 \\  0 & 0 & \sigma_{xz}^x \\ 0 & -\sigma_{xz}^x & 0 \end{pmatrix}$}
        & \multirow{3}{*}{$-$} \\
        $6/mm'm'$  &  &  &  & & \\
        $\bar{3}m'$  &  &  &  & &   
    \end{tabular}
    \end{ruledtabular}
    \label{tab:table}
\end{table*}

(i) Any MLG that contains pure time-reversal $1'$ (reversal of the magnetic texture maps the crystal onto itself modulo a translation) is incompatible with a SCV and a MSHE because $\omega$, $\vec{\varOmega}$, and $\underline{W}$ transform as magnetic multipoles.

(ii) The MLG $m \bar{3} m$ of cubic systems does not allow for a magnetization ($\vec{\varOmega}=\vec{0}$) but for $\omega$ and $W_x^x = W_y^y = W_z^z$, from which
\begin{align*}
     \stackrel{\mathrm{Eq.~\eqref{eq:vorticity-decomp}}}{\to} 
     \omega_x^x = \omega_y^y = \omega_z^z
     &\stackrel{\mathrm{Eq.~\eqref{eq:connectionMSHEvorticity}}}{\to} 
     \sigma_{\mathrm{MSHE},x}^x = \sigma_{\mathrm{MSHE},y}^y = \sigma_{\mathrm{MSHE},z}^z
     \\
     &\stackrel{\mathrm{Eq.~\eqref{eq:MSHEvector}}}{\to}
     \sigma^{x,\mathrm{(a)}}_{yz} = \sigma^{y,\mathrm{(a)}}_{zx} = \sigma^{z,\mathrm{(a)}}_{xy}
\end{align*}
follows. MSHEs with mutually orthogonal spin, flow, and force directions are expected, a situation known from the SHE in nonmagnetic cubic materials. The SHE and MSHE can be disentangled by their opposite time-reversal signature which can be probed by a reversal of the magnetic texture~\cite{Kimata2019}.

(iii) The MLG $4/mm'm'$ admits of a magnetization, $\vec{\varOmega} = (0, 0, \varOmega_z)^\mathrm{T}$, but neither of $\omega$ nor of $\underline{W}$. We find
\begin{align*}
    W_x^y = 0, \; \varOmega_z \ne 0
    &\stackrel{\mathrm{Eq.~\eqref{eq:vorticity-decomp}}}{\to} 
    \omega_x^y = - \omega_y^x
    \stackrel{\mathrm{Eq.~\eqref{eq:connectionMSHEvorticity},\eqref{eq:MSHEvector}}}{\to}
    \sigma^{y,\mathrm{(a)}}_{yz} = - \sigma^{x,\mathrm{(a)}}_{zx}.
\end{align*}
In contrast to the anomalous Hall effect (AHE), a magnetization (along $z$) does not cause transverse transport within a plane perpendicular to it ($xy$ plane), but in planes that contain itself ($xz$ and $yz$ planes). Only the transported spin component has to be normal to the magnetization ($x$ and $y$). Moreover, it has to lie within the plane of transport. Thus, although tetragonal ferromagnets allow both for the AHE and the MSHE, the spin current attributed to the MSHE is a pure spin current because the AHE-induced current flows in a different plane.  This scenario was outlined in Sec.~\ref{sec:overview} via Eq.~\eqref{eq:vorticities-in-ferro}.

\begin{SCfigure*}
    \centering
    \includegraphics[width=0.75\textwidth]{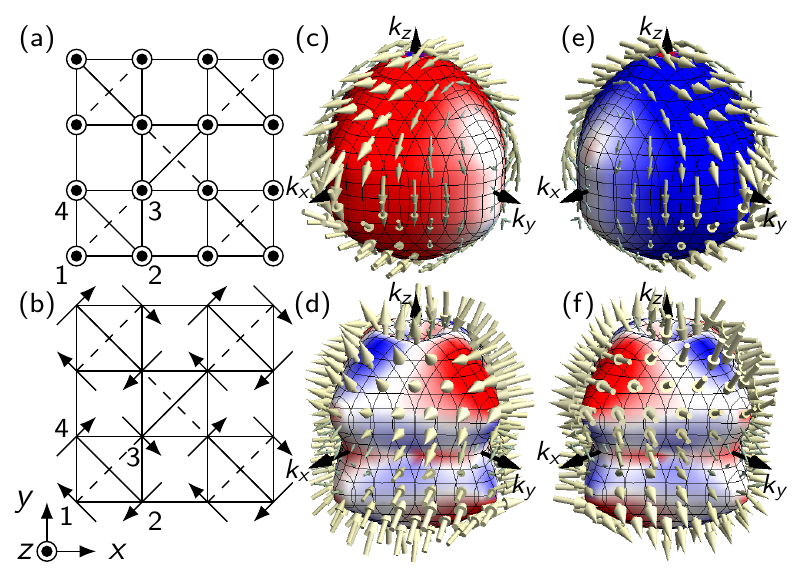}
    \caption{MSHE in the pyrochlore lattice for (a) ferromagnetic (MLG $4/mm'm'$) and (b) antiferromagnetic textures (MLG $4'/mm'm$). The pyrochlore lattice is projected onto the $xy$ plane, such that the tetrahedra appear as squares. (c) and (d) depict Fermi surfaces with arrows indicating $\vec{J}^x_{\vec{k}}$ and the color scale depicts the $y$ component of $\hat{\vec{v}}_{\vec{k}} \times  \vec{J}^x_{\vec{k}}$ (blue/white/red indicates negative/zero/positive values; the value range is symmetric about zero). (e) and (f) as (c) and (d) but for $\vec{J}^y_{\vec{k}}$ and the $x$ component of $\hat{\vec{v}}_{\vec{k}} \times  \vec{J}^y_{\vec{k}}$.}
    \label{fig:Pyro}
\end{SCfigure*}

(iv) A magnetic toroidal quadrupole $\underline{W}$ also allows for the spin transport discussed in (iii). Consider the MLG $4'/mm'm$ that permits only a nonzero $W_x^y$, which translates to $\sigma^{y,\mathrm{(a)}}_{yz} = \sigma^{x,\mathrm{(a)}}_{zx}$. Compared to (iii), only the relation of the signs of nonzero components has changed. For a geometry as depicted in Fig.~\ref{fig:MSHEFerro} (four-fold rotational axis aligned with the $z$ direction), this reversed sign translates into a MSHE spin accumulation with a polarization that alternates between pointing parallel and antiparallel to the surface normal.

(v) The MLG $m'm'm$ combines the scenarios of (iii) and (iv): $\varOmega_z$, $W_x^y$, $\sigma^{y,\mathrm{(a)}}_{yz}$, and $\sigma^{x,\mathrm{(a)}}_{zx}$ may be nonzero but there is no additional symmetry-imposed relation between the latter two.

(vi) An MSHE has been experimentally established in the noncollinear antiferromagnet Mn$_3$Sn~\cite{Kimata2019} that---depending on the spin orientation---belongs either to the MLG $2'/m'$ or to $2/m$ \footnote{If the spin orientation breaks all crystal symmetries, Mn$_3$Sn belongs to the MLG $\bar{1}$, which is, however, unlikely in the absence of a magnetic field, because of anisotropies along high symmetry directions of the lattice.}. These are the same MLGs we shall discuss in the context of a magnetized Rashba electron gas with warping (Sec.~\ref{sec:MSHEinRashba}). Please note that the MLG $2/m$ allows for $\sigma_{xy}^z \not= 0$, that is a MSHE in the $xy$ plane with out-of-plane polarized spins currents (a geometry similar to the conventional SHE), whereas $2'/m'$ does not. Thus, upon rotation of the coplanar magnetic texture of Mn$_3$Sn, one could switch the MLGs and thereby engineer the transport of out-of-plane polarized spins; this effect awaits experimental verification (the experimental setup in Ref.~\onlinecite{Kimata2019} was sensitive to in-plane spin polarizations).

\section{Examples}
\label{sec:examples}
With the above results at hand, we now address selected examples for various MLGs. Section~\ref{sec:minimal-req} focuses on minimal requirements for a MSHE and illustrates its interpretation in terms of spin current vorticities. In Sec.~\ref{sec:MSHEinRashba} we make contact with Rashba materials whose MLGs cover Mn$_3$Sn. Finally, we consider a magnetic spin Nernst effect in insulating materials (Sec.~\ref{sec:MSNE}).

\subsection{Minimal requirements for a MSHE}
\label{sec:minimal-req}
According to the points (iii) and (iv) in Section~\ref{sec:symmetry_analysis}, compatibility of a MLG with either a magnetization (e.\,g., MLG $4/mm'm'$) or a magnetic toroidal quadrupole (e.\,g., MLG $4'/mm'm$) suffices for a MSHE\@. To show explicitly the spin current vortex about the Fermi surface [in the sense of Eq.~\eqref{eq:MSHE_cond}], we consider the $sd$ Hamiltonian
\begin{align}
    H = \sum_{\langle ij \rangle}
      \vec{c}_{i}^\dagger \left( t + \mathrm{i} \alpha \vec{\tau} \cdot \hat{\vec{d}}_{ij} \right) \vec{c}_{j}
      + J \sum_{i} \vec{c}_{i}^\dagger \left( \vec{\tau} \cdot \hat{\vec{m}}_{i} \right) \vec{c}_{i}
      \label{eq:TBHam}
\end{align}
on the pyrochlore lattice [Fig.~\ref{fig:Pyro}(a)] which consists of corner-sharing tetrahedra \cite{Bzdusek2015}. $\vec{c}^\dagger_{i}$ ($\vec{c}_{i}$) creates (annihilates) an electron spinor at site $i$, $\vec{\tau}^\mathrm{T} = (\underline{\tau}^x, \underline{\tau}^y, \underline{\tau}^z)$ is the vector of Pauli matrices. The hopping (with amplitude $t$) of electrons is accompanied by a spin rotation due to SOC (with amplitude $\alpha$). The unit vectors
\begin{equation}
\begin{aligned}
   \hat{\vec{d}}_{12} = \frac{\sqrt{2}}{2}
       \begin{pmatrix}
          -1 \\ 0 \\ 1
       \end{pmatrix}
       , \quad
   \hat{\vec{d}}_{13} = \frac{\sqrt{2}}{2}
       \begin{pmatrix}
          1 \\ -1 \\ 0
       \end{pmatrix}
       , \quad
    \hat{\vec{d}}_{14} = \frac{\sqrt{2}}{2}
       \begin{pmatrix}
          0 \\ 1 \\ -1
       \end{pmatrix}
       , 
     \\
   \hat{\vec{d}}_{23} = \frac{\sqrt{2}}{2}
       \begin{pmatrix}
          0 \\ 1 \\ 1
       \end{pmatrix}
       , \quad
    \hat{\vec{d}}_{24} = \frac{\sqrt{2}}{2}
       \begin{pmatrix}
          -1 \\ -1 \\ 0
       \end{pmatrix}
       , \quad
   \hat{\vec{d}}_{34} = \frac{\sqrt{2}}{2}
       \begin{pmatrix}
          1 \\ 0 \\ 1
       \end{pmatrix}
    \label{eq:dvectors}
\end{aligned}
\end{equation}
specify the directions of the effective SOC\@. Each of the $\hat{\vec{d}}_{ij}$ is orthogonal to the $i$--$j$ bond and lies within a face of a cube that encloses a tetrahedron.  $\hat{\vec{d}}_{ij} = -\hat{\vec{d}}_{ji}$ implies $\sum_{j=1}^4 \hat{\vec{d}}_{ij} = \vec{0}$ for each $i = 1,\ldots,4$.

Via Hund's coupling $J$, the electron spins are connected with the local magnetic moments $\hat{\vec{m}}_{i}$ (black arrows in Fig.~\ref{fig:Pyro}). The ferromagnetic texture in (a), with $\hat{\vec{m}}_{i} = (0,0,1)^\mathrm{T}$, belongs to the MLG $4/mm'm'$, while the antiferromagnetic texture in (b) belongs to $4'/mm'm$ (the notation of the MLGs matches that of the respective magnetic point groups):
\begin{equation}
\begin{aligned}
   \hat{\vec{m}}_{1} = \frac{\sqrt{2}}{2}
       \begin{pmatrix}
          -1 \\ 1 \\ 0
       \end{pmatrix}
       , \quad
   \hat{\vec{m}}_{2} = \frac{\sqrt{2}}{2}
       \begin{pmatrix}
          -1 \\ -1 \\ 0
       \end{pmatrix}
       , \\
    \hat{\vec{m}}_{3} = \frac{\sqrt{2}}{2}
       \begin{pmatrix}
          1 \\ -1 \\ 0
       \end{pmatrix}
       , \quad
   \hat{\vec{m}}_{4} = \frac{\sqrt{2}}{2}
       \begin{pmatrix}
          1 \\ 1 \\ 0
       \end{pmatrix}
       .
\end{aligned}
\end{equation}

The latter MLG $4'/mm'm$ is compatible with a magnetic toroidal quadrupole ($W_x^y \ne 0$), which is evident from the texture itself: the quadruple of spins labelled ${1, 2, 3, 4}$ in Fig.~\ref{fig:Pyro}(b) produces a toroidal dipole moment
\begin{align}
    \vec{T} \propto \sum_{i=1}^4 \vec{r}_i \times \hat{\vec{m}}_i,
\end{align}
in which coordinates $\vec{r}_i$ are taken with respect to the tetrahedron's center of mass. We obtain $\vec{T} = (0,0,T_z)^\mathrm{T}$ with $T_z < 0$, i.\,e., the toroidal dipole points into the paper plane. However, any neighboring tetrahedron features magnetic moments with an opposite circulation direction, giving rise to the opposite toroidal dipole moment ($T_z > 0$). The regular array of alternating up and down-pointing toroidal dipoles causes a zero net toroidal dipole (as expected for $4'/mm'm$ because the Jahn symbol $aV$ of $\vec{T}$) but a nonzero net magnetic toroidal quadrupole \cite{Suzuki2019}.

We have diagonalized the Hamiltonian \eqref{eq:TBHam} in reciprocal space. Since the magnetic unit cell contains the same number of sites as the structural unit cell, there are four electronic bands (not shown). Fig.~\ref{fig:Pyro}(c)--(f) show representative iso-energy cuts (Fermi surfaces) with the arrows depicting $\vec{J}^x_{\vec{k}}$ [in panels (c) and (d); band index suppressed] and $\vec{J}^y_{\vec{k}}$ [in panels (e) and (f)], respectively. The color scales visualize particular integrands in Eq.~\eqref{eq:MSHE_cond}: the $y$ component of $\hat{\vec{v}}_{\vec{k}} \times  \vec{J}^x_{\vec{k}}$ in (c) and (d) and the $x$ component of $\hat{\vec{v}}_{\vec{k}} \times  \vec{J}^y_{\vec{k}}$ in (e) and (f). The spin current is defined as usual by $\underline{J}^\gamma_{\vec{k},\mu} = \frac{1}{2} \{ \underline{s}^\gamma , \underline{v}_{\vec{k},\mu} \}$ ($\underline{s}^\gamma$ $\gamma$-spin operator).

The spin current circulation about the Fermi surface, that is the spin current vortex, is clearly identified in panels (c) and (e). The respective integrals
\begin{align}
    \sigma_{zx}^{x,\mathrm{(a)}} \propto  \sum_{n} \oint_{\varepsilon_n = \varepsilon_\mathrm{F}} \left.  \hat{\vec{v}}_{n \vec{k}} \times \vec{J}^x_{n \vec{k}} \right|_{y} \, \mathrm{d} S
\end{align}
and
\begin{align}
    \sigma_{yz}^{y,\mathrm{(a)}} \propto  \sum_{n} \oint_{\varepsilon_n = \varepsilon_\mathrm{F}} \left.  \hat{\vec{v}}_{n\vec{k}} \times \vec{J}^y_{n \vec{k}} \right|_{x} \, \mathrm{d} S
\end{align}
are nonzero [either red (c) or blue color (e) dominates]. Due to the symmetry of the value range, one finds 
\begin{align}
    -\sigma_{zx}^{x,\mathrm{(a)}} = \sigma_{xz}^{x,\mathrm{(a)}} = \sigma_{yz}^{y,\mathrm{(a)}},
\end{align}
as was confirmed numerically. These findings agree fully with point (iii) of Sec.~\ref{sec:symmetry_analysis} and with the $4/mm'm'$ row of Table~\ref{tab:table}.

A similar, albeit less striking observation can be made for panels (d) and (f). Red dominates slightly over blue in both cases, visualizing nonzero integrals for $\sigma_{zx}^x$ and $\sigma_{yz}^y$. In accordance with point (iv) of Sec.~\ref{sec:symmetry_analysis} and the entry for $4'/mm'm$ in Table~\ref{tab:table},
\begin{align}
    \sigma_{zx}^{x,\mathrm{(a)}} = -\sigma_{xz}^{x,\mathrm{(a)}} = \sigma_{yz}^{y,\mathrm{(a)}}
\end{align}
holds.

For all other components of $\hat{\vec{v}}_{\vec{k}} \times  \vec{J}^\gamma_{\vec{k}}$ that are not shown in Fig.~\ref{fig:Pyro}, the color distribution on the Fermi surface---blue and red appear equally---indicate magnetic spin Hall conductivities of zero, in agreement with Table~\ref{tab:table}.

\subsection{MSHE in Rashba materials}
\label{sec:MSHEinRashba}
For the three-dimensional models addressed in the preceding section, we concentrated on the spin current circulation about the Fermi surface. Similar conclusions can be drawn from calculated SCVs which one could represent as `vortex lines' of the field $\vec{J}^\gamma_{\vec{k}}$. Since this makes for hardly interpretable three-dimensional pictures, we focus now on a two-dimensional model for which the SCV is clearly identified.

\begin{figure}
    \centering
    \includegraphics[width=1.0\columnwidth]{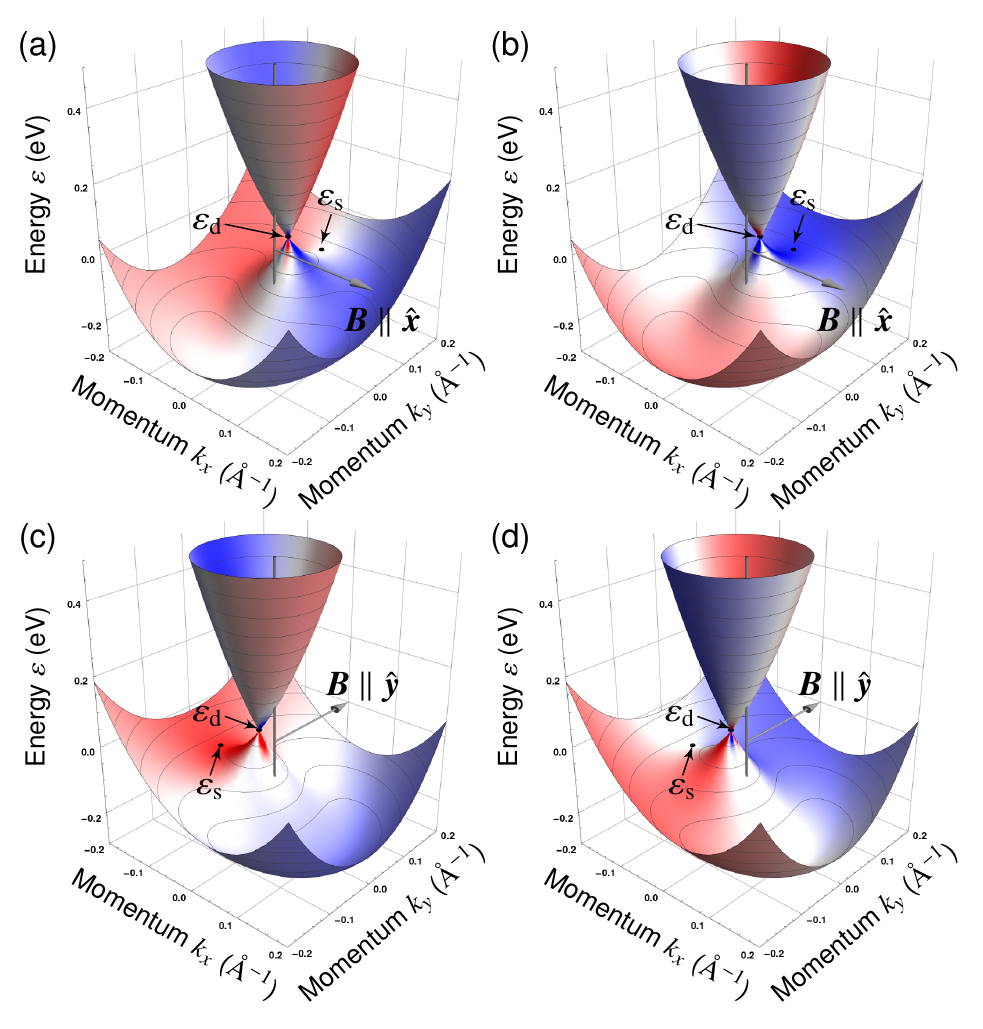}
    \caption{Two-dimensional electron gas with Rashba SOC and in-plane magnetic field along $x$ [(a) and (b)] or $y$ [(c), (d)].  $\varepsilon_\mathrm{d}$ and $\varepsilon_\mathrm{s}$ are the energy of the degeneracy point and of the saddle point. The color scales represent spin current vorticities $\omega^x_{\vec{k},z}$ [(a), (c)] and $\omega^y_{\vec{k},z}$ [(b), (d); blue/white/red color indicates negative/zero/positive values]. For details, see text.}
    \label{fig:Rashba}
\end{figure}

\begin{SCfigure*}
    \centering
    \includegraphics[scale=0.9]{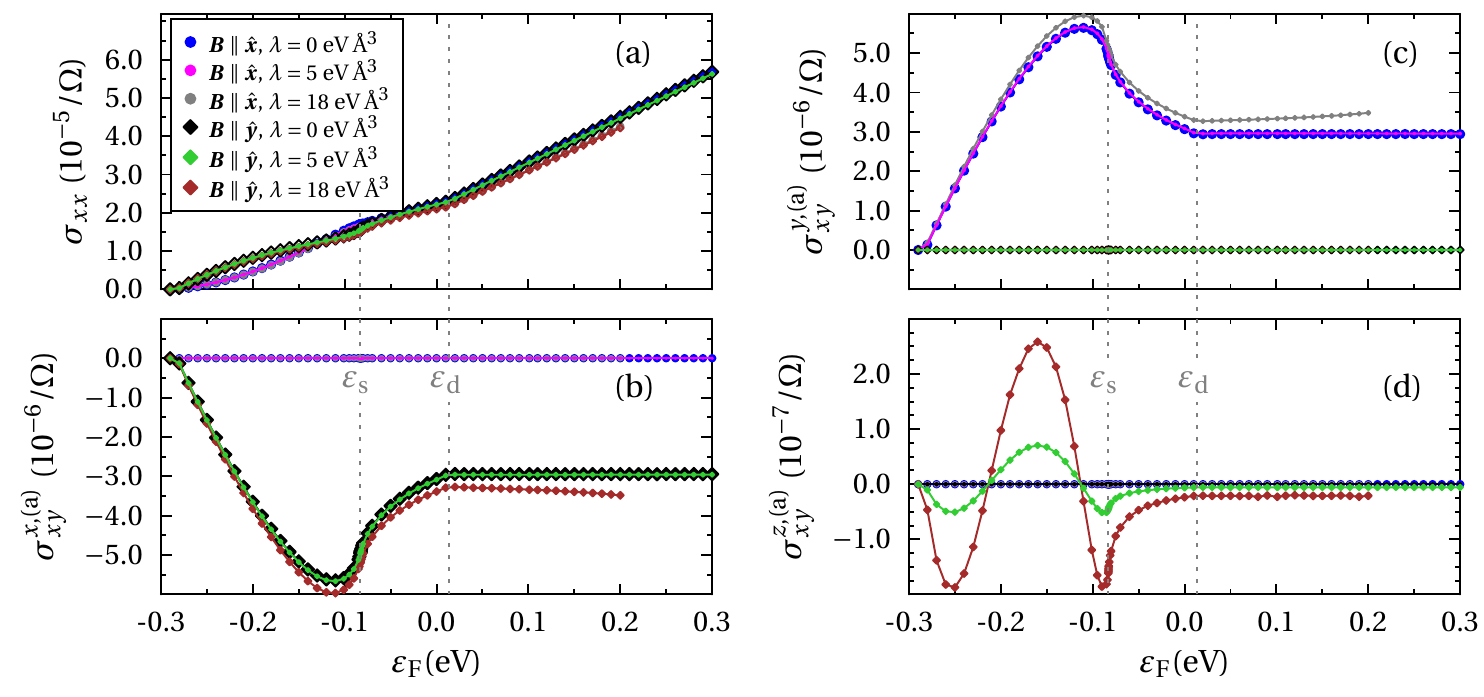}
    \caption{Charge conductivity [(a)] and MSHE [(b)-(d)] in a Rashba system with in-plane magnetic field $\vec{B}$ and hexagonal warping. Model parameters as for Fig.~\ref{fig:Rashba}, except for the warping strength $\lambda = \SI{18}{\electronvolt\angstrom\tothe{3}}$. For $\lambda = \SI{18}{\electronvolt \angstrom \tothe{3}}$, the model is applicable only for $\varepsilon< \SI{0.2}{\electronvolt}$.  For comparison, systems without ($\lambda = 0$) and with reduced warping ($\lambda = \SI{5}{\electronvolt\angstrom\tothe{3}}$) are considered. Circles (squares) for $\vec{B} \parallel \hat{\vec{x}}$ ($\vec{B} \parallel \hat{\vec{y}}$); $\underline{\sigma}^{\gamma,\mathrm{(a)}}$ multiplied by $\frac{2 e}{\hbar}$ to better compare with the charge conductivity.}
    \label{fig:Rashba_MSHE}
\end{SCfigure*}

\subsubsection{Magnetized Rashba model and its band structure}
Recent progress on magnetism in two-dimensional electron gases motivates to demonstrate the existence of SCVs in an in-plane magnetized Rashba model with Hamiltonian
\begin{align}
    \underline{H} &= \underbrace{\frac{\hbar^2 k^2}{2m} + \alpha_\mathrm{R} \left(k_x \underline{\tau}^y - k_y \underline{\tau}^x \right)}_{\underline{H}_\mathrm{R}} + \underbrace{\mu_\mathrm{B}\vec{B} \cdot \vec{\tau}}_{\underline{H}_\mathrm{Zee}}, \label{eq:RashbaHam}
\end{align}
($\underline{\tau}^i$ Pauli matrices, $m$ effective mass, and $\alpha_\mathrm{R}$ Rashba parameter); hexagonal warping is accounted for later. The continuous rotational symmetry of $\underline{H}_\mathrm{R}$ is broken by an in-plane exchange field $\vec{B} = (B_x, B_y, 0)^\mathrm{T}$ ($\mu_\mathrm{B}$ Bohr's magneton). The set of parameters ($m = 0.32 \, m_\mathrm{e}$, $\alpha_\mathrm{R} = 2.95\, \si{\electronvolt\angstrom}$, with $m_\mathrm{e}$ electron mass) corresponds to those of the ordered $(\sqrt{3} \times \sqrt{3})R30^\circ$ Bi/Ag$(111)$ surface alloy \cite{Ast2007, Meier2008, Bentmann_2009, Frantzeskakis2011}; we set $\mub B = \unit[0.1]{eV}$.

For $\vec{B}$ in $x$ direction, the two bands are degenerate  at a point on the $k_y$ axis [panels (a) and (b) of Fig.~\ref{fig:Rashba}; at energy $\varepsilon_\mathrm{d}$]; likewise for $\vec{B}$ along $y$, the bands are degenerate at a point on the $k_x$ axis [(c) and (d)]. On top of that, the lower band has a saddle point at energy $\varepsilon_\mathrm{s}$. Overall, the band structure merely exhibits a $k_x \to -k_x$ [(a), (b)] or  a $k_y \to -k_y$ symmetry [(c), (d)], rendering iso-energy lines anisotropic.

\subsubsection{Symmetries, spin current vorticities, and spin conductivity}
With $\vec{B}$ in $x$ direction ($y$ direction), the model shows nonzero $\varOmega_x$ ($\varOmega_y$) and, thus, nonzero $\sigma^{y, \mathrm{(a)}}_{xy}$ ($\sigma^{x, \mathrm{(a)}}_{xy}$). Recall that transport takes place in a plane containing the magnetization and that the transported spin component is orthogonal to the magnetization. Since we focus on transport in the $xy$ plane, we consider neither $\sigma^{z, \mathrm{(a)}}_{xz}$  nor $\sigma^{z, \mathrm{(a)}}_{yz}$ although allowed by $\varOmega_x$ or $\varOmega_y$.

The momentum-resolved SCVs, shown by color in Fig.~\ref{fig:Rashba}, exhibit a band antisymmetry, $\omega^x_{1,\vec{k},z} = -\omega^x_{2,\vec{k},z}$ and $\omega^y_{1,\vec{k},z} = - \omega^y_{2,\vec{k},z}$ ($1$ and $2$ band indices), which is a feature of a two-band model. Moreover, the SCVs exhibit the following reflection (anti-)symmetries:
\begin{subequations}
\begin{align}
    \vec{B} &\parallel \hat{\vec{x}}: \quad \omega^x_{n,k_x,k_y,z} = -\omega^x_{n,-k_x,k_y,z}, \quad \omega^y_{n,k_x,k_y,z} = \omega^y_{n,-k_x,k_y,z},
    \\
    \vec{B} &\parallel \hat{\vec{y}}: \quad \omega^x_{n,k_x,k_y,z} = \omega^x_{n,k_x,-k_y,z}, \quad \omega^y_{n,k_x,k_y,z} = -\omega^y_{n,k_x,-k_y,z}.
\end{align}
\end{subequations}
Even without explicit calculations one verifies that any Fermi sea integral over $\omega^x_{\vec{k},z}$ for $\vec{B} \parallel \hat{\vec{x}}$ [panel (a)] or $\omega^y_{\vec{k},z}$ for $\vec{B} \parallel \hat{\vec{y}}$ [panel (d)] equals zero due to these antisymmetries. Equation~\eqref{eq:MSHE_cond_vorticity} gives $\sigma^{x, \mathrm{(a)}}_{xy} = 0$ for $\vec{B} \parallel \hat{\vec{x}}$ and $\sigma^{y, \mathrm{(a)}}_{xy} = 0$ for $\vec{B} \parallel \hat{\vec{y}}$. In contrast, the integrals over $\omega^y_{\vec{k},z}$ for $\vec{B} \parallel \hat{\vec{x}}$ (b) or $\omega^x_{\vec{k},z}$ for $\vec{B} \parallel \hat{\vec{x}}$ (c) are nonzero, as becomes especially plausible for low energies, at which only red (b) or blue (c) shows up.

For a quantitative analysis, we address the energy dependence of the magnetic spin Hall conductivity. First, we recall that of the charge conductivity $\sigma_{xx}$, as shown in Fig.~\ref{fig:Rashba_MSHE}(a). For low Fermi energies $\varepsilon_\mathrm{F}$, the direction of the magnetic field strongly affects the shape of the iso-energy lines, leading to different $\sigma_{xx}$ for $\vec{B} \parallel \hat{\vec{x}}$ and $\vec{B} \parallel \hat{\vec{y}}$, respectively [compare blue and black symbols in Fig.~\ref{fig:Rashba_MSHE}(a)]. At higher energies SOC dominates over exchange and, thus, the shape of the iso-energy lines depends marginally on the direction of $\vec{B}$: there is barely a difference in $\sigma_{xx}$ for $\vec{B}\parallel \hat{\boldsymbol{x}}$ and $\vec{B}\parallel \hat{\boldsymbol{y}}$. 

\begin{figure}
    \centering
    \includegraphics[width=\columnwidth]{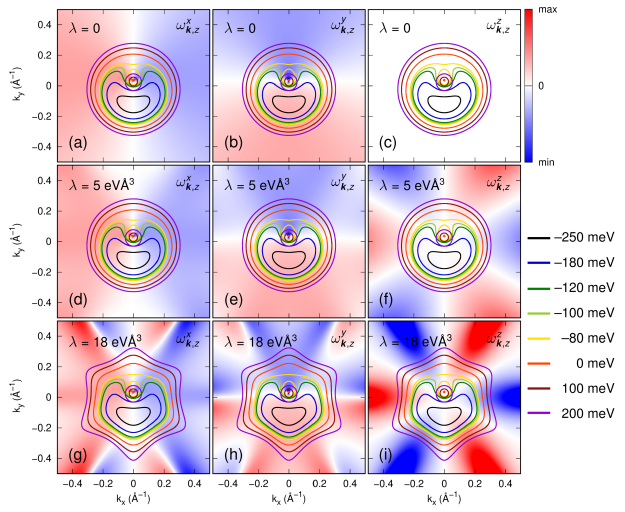}
    \caption{Spin current vorticity (red/blue color scale) and iso-energy lines (colored lines) of the lower band of Rashba systems with and without hexagonal warping in the presence of an in-plane magnetic field $\vec{B} \parallel \hat{\vec{x}}$. Model parameters as in Fig.~\ref{fig:Rashba_MSHE}.}
    \label{fig:vorticity_Bx}
\end{figure}

\begin{figure}
    \centering
    \includegraphics[width=\columnwidth]{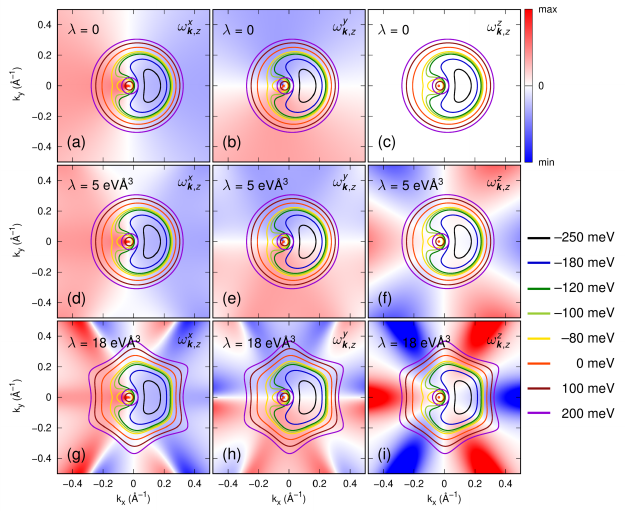}
    \caption{As Fig.~\ref{fig:vorticity_Bx} but for $\vec{B} \parallel \hat{\vec{y}}$.}
    \label{fig:vorticity_By}
\end{figure}

For understanding better the energy dependence of the MSHE depicted in Figs.~\ref{fig:Rashba_MSHE}(b)--(d), we inspect the SCVs $\omega_{\vec{k},z}^x$, $\omega_{\vec{k},z}^y$, and $\omega_{\vec{k},z}^z$ for $\vec{B} \parallel \hat{\boldsymbol{x}}$ and $\vec{B} \parallel \hat{\boldsymbol{y}}$, respectively (Figs.~\ref{fig:vorticity_Bx} and \ref{fig:vorticity_By}). As expected, $\sigma_{xy}^{x,\mathrm{(a)}}$ ($\sigma_{xy}^{y,\mathrm{(a)}}$) vanishes for $B_y=0$ ($B_x=0$), which is illustrated in Fig.~\ref{fig:vorticity_Bx}(a) [Fig.~\ref{fig:vorticity_By}(b)]: the contributions from Fermi sea regions with positive and negative SCV cancel. In contrast, a finite conductivity $\sigma_{xy}^{y,\mathrm{(a)}}$ ($\sigma_{xy}^{x,\mathrm{(a)}}$) occurs due to incomplete cancellation [Figs.~\ref{fig:vorticity_Bx}(b) and \ref{fig:vorticity_By}(a)]. 

As sketched in Fig.~\ref{fig:Rashba}, the bands are degenerate at $\varepsilon_\mathrm{d}\approx \SI{8}{\milli\electronvolt}$. For Fermi levels below the degeneracy, only the lower band is occupied. Increasing the Fermi level from the band edge, the absolute value of $\sigma_{xy}^{y,\mathrm{(a)}}$ ($\sigma_{xy}^{x,\mathrm{(a)}}$) increases [cf.~Figs.~\ref{fig:Rashba_MSHE}(c) and (b)] due to the growing number of states contributing to the MSHE [cf.~black, blue, and green Fermi lines in Figs.~\ref{fig:vorticity_Bx} or \ref{fig:vorticity_By}]. 

Around the saddle point at $\varepsilon_\mathrm{s}\approx -\SI{83}{\milli\electronvolt}$ the SCV changes sign and the states contribute oppositely to the magnetic spin Hall conductivity, leading to an extremum of $\sigma_{xy}^{y,\mathrm{(a)}}$ ($\sigma_{xy}^{x,\mathrm{(a)}}$) close to $\varepsilon_\mathrm{s}$. 

Above $\varepsilon_\mathrm{d}$, both bands are occupied (in Figs.~\ref{fig:vorticity_Bx} and \ref{fig:vorticity_By}, the inner iso-energy line corresponds to the upper band, the outer to the lower band). Only the SCV of the lower band is shown, but the upper band's SCV differs from the lower band's only by sign. Thus, regions in $\vec{k}$ space in which both bands are occupied do not contribute to $\sigma_{xy}^{\gamma,\mathrm{(a)}}$.

With increasing $\varepsilon_\mathrm{F}$ the additional contribution of the lower band's states to the MSHE is compensated by the states in the upper band, thus,  $\sigma_{xy}^{y,\mathrm{(a)}}$ and $\sigma_{xy}^{x,\mathrm{(a)}}$ are almost independent of $\varepsilon_\mathrm{F}$.

The magnetic spin Hall angle, defined as $\alpha^\gamma_\mathrm{SH} =\frac{2e}{\hbar} \times 2 \sigma^{\gamma,\mathrm{(a)}}_{xy} / (\sigma_{xx}+ \sigma_{yy} )$ is sizable (up to $70\%$) near the band edge and decreases for larger $\varepsilon_\mathrm{F}$. Above $\varepsilon_\mathrm{d}$, it is in the order of $10\%$, which may be considered substantial. Its energy dependence is dominated by the almost linear energy dependence of the charge conductivity.

\subsubsection{Effect of hexagonal warping}
To come closer to realistic materials, a hexagonal warping term
\begin{align}
    \underline{H}_\mathrm{w} = \frac{\mathrm{i} \lambda}{2} \left( k_{+}^3 - k_{-}^3 \right) \underline{\tau}^z,  \quad k_\pm = k_x \pm \mathrm{i} k_y, \label{eq:warping}
\end{align}
is added to the Hamiltonian ~\eqref{eq:RashbaHam}. For $(\sqrt{3} \times \sqrt{3})R30^\circ$ Bi/Ag$(111)$, the strength of the warping is $\lambda= \SI{18}{\electronvolt\angstrom\tothe{3}}$~\cite{Frantzeskakis2011}. Similar to the exchange field, $\underline{H}_\mathrm{w}$ breaks the continuous rotation symmetry of $\underline{H}_\mathrm{R}$, leaving the $xz$ plane as a mirror plane. 

For $\vec{B} \parallel \hat{\vec{y}}$ this mirror plane is retained (MLG $2/m$; Table~\ref{tab:table}). Since $\underline{H}_\mathrm{w}$ introduces a spin-$z$ component [notice $\underline{\tau}^z$ in Eq.~\eqref{eq:warping}], we expect nonzero $\sigma_{xy}^{z,\mathrm{(a)}}$ [associated with $W_z^z$, as explained in (ii); cf.\  Fig.~\ref{fig:Rashba_MSHE}(b) and (d)]. For weak warping ($\lambda = \SI{5}{\electronvolt \angstrom \tothe{3}}$), the band structure as well as the SCV $\omega_{\vec{k},z}^x$ appear mildly affected by the additional SOC [Fig.~\ref{fig:vorticity_By}(d)]. Thus, $\sigma_{xy}^{x,\mathrm{(a)}}$ is weakly influenced by warping.  However, a finite spin current vorticity $\omega_{\vec{k},z}^z$ gives rise to a nonzero $\sigma_{xy}^{z,\mathrm{(a)}}$. The sign changes of $\sigma_{xy}^{z,\mathrm{(a)}}$ for $\varepsilon_\mathrm{F}<\varepsilon_\mathrm{s}$ are due the anisotropy of the Fermi lines and the alternating sign of $\omega_{\vec{k},z}^z$ in reciprocal space [Fig.~\ref{fig:vorticity_By}(f) and (i)]. 

For stronger warping ($\lambda = \SI{18}{\electronvolt \angstrom \tothe{3}}$), the energy dispersion and $\omega_{\vec{k},z}^x$ are remarkably modified, which leads to a slightly enhanced MSH conductivity $\sigma_{xy}^{x,\mathrm{(a)}}$ at any $\varepsilon_\text{F}$. Furthermore, the absolute value of $\sigma_{xy}^{z,\mathrm{(a)}}$ is increased.

$\vec{B} \parallel \hat{\vec{x}}$ makes the $xz$ plane a time-reversal mirror plane (MLG $2'/m'$; Table~\ref{tab:table}). Instead of $\sigma_{xy}^{x,\mathrm{(a)}}$ and $\sigma_{xy}^{z,\mathrm{(a)}}$, only $\sigma_{xy}^{y,\mathrm{(a)}}$ is now nonzero [Fig.~\ref{fig:Rashba_MSHE}(b)], and its energy dependence is equivalent to that of $\sigma_{xy}^{x,\mathrm{(a)}}$ for $\vec{B}\parallel \hat{\vec{y}}$. Note in particular that Figs.~\ref{fig:vorticity_Bx} (f) and (i) demonstrate that the Fermi sea integral over $\omega_{\vec k,z}^z$ vanishes for symmetry reasons.

\subsubsection{Concluding remarks and applicability to real materials}
To conclude, in-plane magnetized Rashba 2DEGs exhibit a MSHE with in-plane polarized spin current ($\sigma^{x,(a)}_{xy}$ and $\sigma^{y,(a)}_{xy}$) if warping is negligibly small. By rotating the in-plane $\vec{B}$ field, the transported spin components can be manipulated. On top of that, the interplay of warping and the direction of $\vec{B}$ allows for nonzero $\sigma^{z,(a)}_{xy}$, thereby causing out-of-plane polarized spin accumulations at the edges of the sample (similar to the conventional SHE). Upon continuous rotation of $\vec{B}$, the magnetic spin Hall conductivity $\sigma^{z,(a)}_{xy}$ alternates from positive ($\vec{B} \parallel \hat{\vec{y}}$) via zero ($\vec{B} \parallel \pm \hat{\vec{x}}$) to negative ($\vec{B} \parallel -\hat{\vec{y}}$) values.

A similar effect is expected for the warped topological Dirac surface states in Sn-doped Bi$_2$Te$_3$ \cite{Fu2009}; the exchange field could be induced by proximity to a ferromagnetic insulator \cite{Eremeev2013}, a strategy giving rise to an AHE \cite{Mogi2019}. Another example is the noncollinear antiferromagnet Mn$_3$Sn with spin textures as shown in Figs.~\ref{fig:nvc}(a) and (d). While the manipulation of in-plane polarized MSHE spin currents by an in-plane field has been successfully demonstrated in Ref.~\onlinecite{Kimata2019}, the manipulation of the out-of-plane polarized MSHE awaits its experimental confirmation.

The Rashba model for a 2DEG is easily extended to three dimensions, in order to cover multiferroic Rashba semiconductors with bulk Rashba SOC, an example being like (GeMn)Te  \cite{Krempaski2016, Krempaski2018, Yoshimi2018}. In equilibrium, the magnetization of (GeMn)Te is parallel to the direction of the ferroelectric polarization \cite{Krempaski2016}; it is conceivable that an in-plane field causes considerable in-plane canting and a MSHE\@.

\subsection{Magnetic spin Nernst effect}
\label{sec:MSNE}

\begin{SCfigure*}
    \centering
    \includegraphics{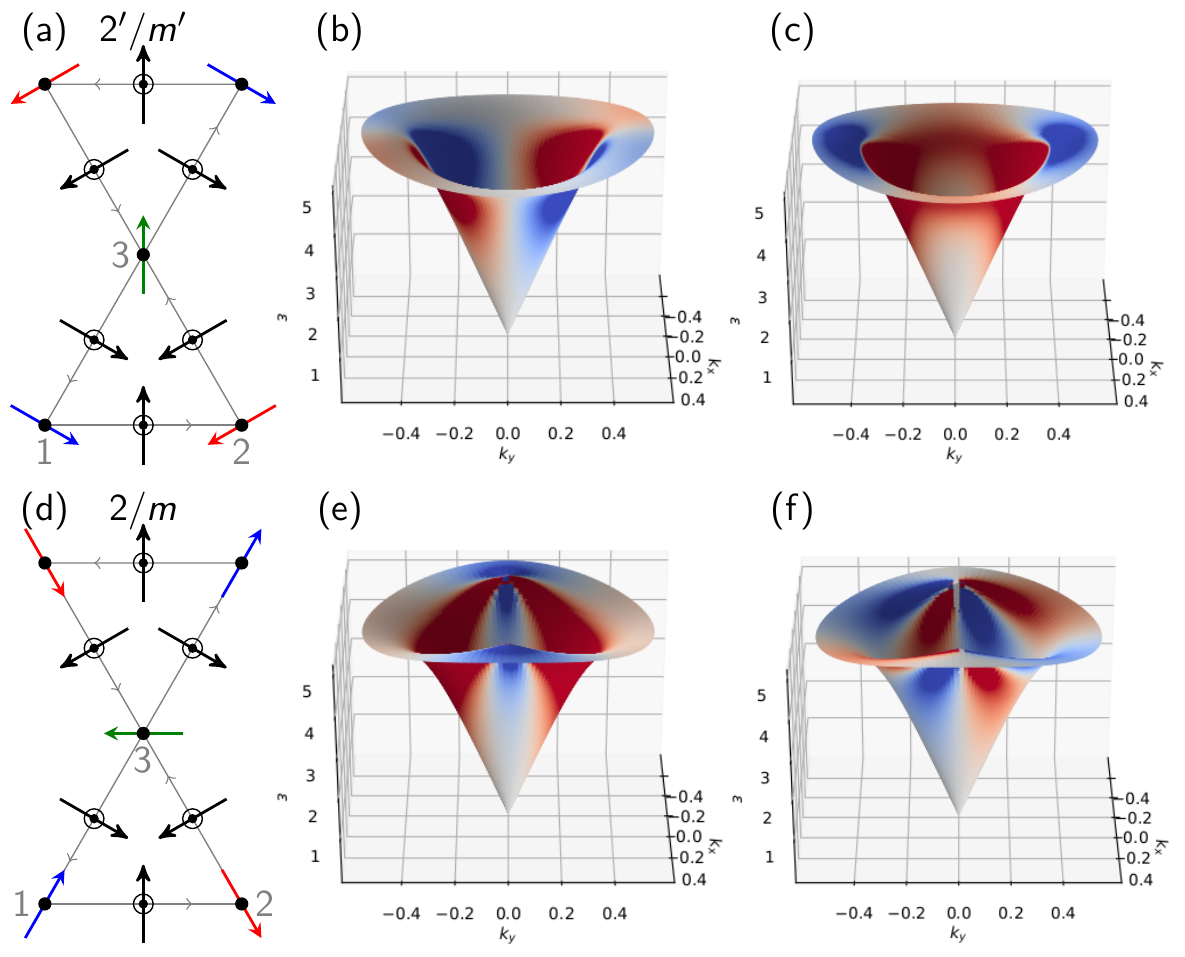}
    \caption{Magnon SCVs for two antiferromagnetic textures on the kagome lattice. Top row: MLG $2'/m'$. (a) Colored arrows at the vertices and black arrows at the bond centers indicate the spin texture and the DMI vectors for counter-clockwise circulation, respectively. (b) and (c) Dispersion relation of the lowest band in the vicinity of the Brillouin zone center. Color scales indicate the value of the SCVs (b) $\omega^x_{\vec{k},z}$ and (c) $\omega^y_{\vec{k},z}$  (red/white/blue stands for positive/zero/negative SCV). Bottom row: as top row but for the MLG $2/m$. (e) and (f) depict $\omega^x_{\vec{k},z}$ and $\omega^y_{\vec{k},z}$, respectively.}
    \label{fig:nvc}
\end{SCfigure*}

The symmetry considerations of Sec.~\ref{sec:symmetry_analysis} apply also to the magnetic spin Nernst effect (MSNE) $\langle \vec{j}^\gamma \rangle = \underline{\alpha}^{\gamma}\, (-\vec{\nabla} T)$ ($\vec{\nabla} T$ temperature gradient) which is determined by the antisymmetric part of the magnetothermal conductivity $\underline{\alpha}^\gamma$. Within linear-response theory \begin{equation}
\begin{aligned}
    \alpha^\gamma_{\mu\nu}(\varpi) = & \frac{1}{TV} \int_0^\infty \mathrm{d} t\, \mathrm{e}^{\mathrm{i}\varpi t} \int_0^\beta \mathrm{d}\kappa \left\langle Q_\nu J^\gamma_\mu(t+\mathrm{i} \hbar \kappa) \right\rangle
    +\tilde{\alpha}^\gamma_{\mu\nu}(\varpi) \label{eq:fullKubo-MSNE}
\end{aligned}
\end{equation}
($Q_\nu$ total heat current) \cite{Han2016ax}. $\tilde{\alpha}^\gamma_{\mu\nu}(\varpi)$ accounts for circulating equilibrium currents that do not contribute to transport \cite{Qin2011}. As far as the intraband contribution is concerned, we write
\begin{align}
    \alpha^{\gamma,\mathrm{odd}}_{\mu\nu,\mathrm{intra}} 
        &= \frac{1}{\varGamma T V} \sum_{n, \vec{k}}
        J^\gamma_{n \vec{k},\mu} Q_{n \vec{k},\nu} 
        \left( - \frac{\partial f_{n \vec{k}}}{\partial \varepsilon} \right)
        \label{eq:spin_Nernst_cond_odd_intra}
\end{align}
and derive the MSNE vector
\begin{align}
    \vec{\alpha}^{\gamma}_\mathrm{MSNE} 
    &\equiv
    \frac{1}{2\varGamma T V} \sum_{n, \vec{k}} \vec{J}^\gamma_{n \vec{k}} \times \vec{Q}_{n \vec{k}}  \left( - \frac{\partial f_{n \vec{k}}}{\partial \varepsilon} \right)
    \nonumber 
    \\
    &=
    -\frac{1}{2\hbar \varGamma T (2\pi)^3} \sum_{n} \int_{-\infty}^{\infty} \mathrm{d} \varepsilon \left( \varepsilon - \varepsilon_\mathrm{F} \right) \vec{\omega}^\gamma(\varepsilon) \left( - \frac{\partial f_{n \vec{k}}}{\partial \varepsilon} \right)
    \label{eq:MSNEvector}
\end{align}
using $\vec{Q}_{n \vec{k}} = (\varepsilon_{n\vec{k}} - \varepsilon_\mathrm{F})\, \vec{v}_{n\vec{k}}$. The SCV $\vec{\omega}^\gamma(\varepsilon)$ at energy $\varepsilon$ is obtained from Eq.~\eqref{eq:vorticity-definition} by replacing $\varepsilon_\mathrm{F}$ by $\varepsilon$. Overall, $\vec{\alpha}^{\gamma}_\mathrm{MSNE}$ and $\vec{\sigma}^{\gamma}_\mathrm{MSHE}$ obey the Mott relation
\begin{align}
    \vec{\alpha}^{\gamma}_\mathrm{MSNE} 
    & = 
    -\frac{1}{T}
    \int_{-\infty}^{\infty} \mathrm{d} \varepsilon \left(- \frac{\partial f_{n \vec{k}}}{\partial \varepsilon} \right) \frac{\varepsilon - \mu}{e} \vec{\sigma}^{\gamma}_\mathrm{MSHE}(T=0,\varepsilon) 
    .
    \label{eq:Mott-relation}
\end{align}
Thus, the symmetry restrictions on $\vec{\sigma}^{\gamma}_\mathrm{MSHE}$ also apply to $\vec{\alpha}^{\gamma}_\mathrm{MSNE}$ and, in particular, the MSNE is also related to the SCV\@. Consequently, a nonzero MSHE implies a nonzero MSNE\@.

We now demonstrate the MSNE in magnetic insulators in which $\nabla_\nu T$ causes  magnonic spin currents. Although the Fermi-Dirac distribution $f_{n\vec{k}}$ in Eq.~\eqref{eq:Mott-relation} has to be replaced by the Bose-Einstein distribution $\rho_{n\vec{k}} = (\mathrm{e}^{\beta \varepsilon_{n \vec{k}}}-1)^{-1}$ and the charge current has to be replaced by the particle current, the connection to the SCV remains. Thus, our aim is to show explicitly the existence of a magnonic SCV\@.

Inspired by the antiferromagnetic magnetic texture of the kagome-lattice compound Mn$_3$Sn---for which a MSHE was demonstrated in Ref.~\onlinecite{Kimata2019}---we consider the spin-wave excitations of this texture. Assuming that the kagome plane is not a mirror plane of a surrounding crystal, a minimal spin Hamiltonian reads \cite{Elhajal2002}
\begin{align}
    H = \frac{1}{2\hbar^2} \sum_{\langle i,j \rangle} \left( 
        J \vec{S}_i \cdot \vec{S}_j 
        + D_\parallel \hat{\vec{d}}_{ij} \cdot \vec{S}_i \times \vec{S}_j
        + D_z \hat{\vec{z}}_{ij} \cdot \vec{S}_i \times \vec{S}_j
        \right).
\end{align}
$J > 0$ and $D_\parallel$ parametrize the antiferromagnetic exchange and the in-plane Dzyaloshinskii-Moriya interaction [DMI; the unit vectors $\hat{\vec{d}}_{ij}$ are depicted in Fig.~\ref{fig:nvc}(a)], respectively. $D_z > 0$ is the strength of out-of-plane DMI, with $\hat{\vec{z}}_{ij} = \pm \hat{\vec{z}}$; the upper (lower) sign is for (anti-)cyclic indices $ij$.

The out-of-plane DMI $D_z > 0$ prefers the coplanar magnetic texture of Fig.~\ref{fig:nvc}(a) as the classical ground state over the coplanar all-in--all-out configuration (our sign convention is opposite to that in Ref.~\onlinecite{Elhajal2002}). The corresponding classical ground state energy is independent of $D_\parallel$ as long as $|D_\parallel / D_z|$ is smaller than a critical value (otherwise a canted all-in--all-out texture becomes the energetic minimum \cite{Elhajal2002}). The irrelevance of $D_\parallel$ as far as the classical energy is concerned imposes an accidental rotational degeneracy: the spins can be rotated about the $z$ axis without a classical energy penalty, in particular by $\uppi/2$ [Fig.~\ref{fig:nvc}(d)]. Since the two textures belong to different magnetic point groups---(a) $2'/m'$ and (d) $2/m$---there seems to be a `classical' ambiguity concerning spin transport. This ambiguity is lifted upon performing linear spin-wave theory about the two classical magnetic ground states. Following Ref.~\onlinecite{Mook2019}, we find that the order-by-quantum-disorder mechanism (harmonic zero-point fluctuations contribute to the ground state energy) prefers texture (a) over (d). Nonetheless, it is instructive to study the SCV for both textures to appreciate the effect of magnetic point group symmetries (details of the linear spin wave theory calculation are given in Appendix~\ref{app:LSWA}).

We concentrate on low energies because these are most relevant when accounting for thermal occupation (Bose-Einstein distribution). The SCVs $\omega^x_{\vec{k},z}$ and $\omega^y_{\vec{k},z}$ of the lowest magnon band in the vicinity of the Brillouin zone center are given for the MLG $2'/m'$ in Fig.~\ref{fig:nvc}(b) and (e) as well as for the MLG $2/m$ in (c) and (f).

Inspection of panels (b) and (f) tells that for each state $\vec{k}$ there is a state $\vec{k}'$ with the same energy but with opposite SCV ($\varepsilon_{1, \vec{k}} = \varepsilon_{1, \vec{k}'}$, $\omega^\gamma_{1, \vec{k},z} = -\omega^\gamma_{1, \vec{k}',z}$), a symmetry also found in the Rashba model (Sec.~\ref{sec:MSHEinRashba}). Irrespective of the distribution function, the local contributions to the integrated $\omega_z^\gamma(\varepsilon)$ [Eq.~\eqref{eq:MSNEvector}] cancel out: $\alpha_{xy}^{x,(a)} = 0$ for the $2'/m'$ texture and $\alpha_{xy}^{y,(a)} = 0$ for the $2/m$ texture. In contrast, the SCVs in Fig.~\ref{fig:nvc}(c) and (e) do not exhibit such an antisymmetry and thus have nonzero integral: $\alpha_{xy}^{y,(a)} \ne 0$ for the $2'/m'$ texture and $\alpha_{xy}^{x,(a)} \ne 0$ for the $2/m$ texture. These findings agree with the $2 \times 2$ $xy$ subtensors of $\underline{\sigma}^{x,(a)}$ and $\underline{\sigma}^{y,(a)}$ for both MLGs (Table~\ref{tab:table}),
\begin{subequations}
\begin{align}
    \underline{\alpha}^{x,(a)}_{2'/m'} =
        \begin{pmatrix}
            0 & 0\\
            0 & 0
        \end{pmatrix} \qq{and}
    \underline{\alpha}^{y,(a)}_{2'/m'} =
        \begin{pmatrix}
            0 & \alpha^{y}_{xy}\\
            -\alpha^{y}_{xy} & 0
        \end{pmatrix},
    \\
    \underline{\alpha}^{x,(a)}_{2/m} =
        \begin{pmatrix}
            0 & \alpha^{x}_{xy}\\
            -\alpha^{x}_{xy} & 0
        \end{pmatrix} \qq{and}
    \underline{\alpha}^{y,(a)}_{2/m} =
        \begin{pmatrix}
            0 & 0\\
            0 & 0
        \end{pmatrix}.
\end{align}
\end{subequations}

One could expect a finite $\alpha^z_{xy}$ for $2/m$, as is the case for the electronic Rashba model studied in Sec.~\ref{sec:MSHEinRashba}. For the texture has no out-of-plane component and the magnon spin is defined with respect to the spin directions offered by the ground state texture (Ref.~\onlinecite{Okuma2017} and  Appendix~\ref{app:LSWA}), this component cannot be captured within the present framework. One may regard this a shortcoming of the definition of magnon spin that has to be treated in the future.

\section{Discussion}
\label{sec:Discussion}
As known from the Barnett effect \cite{Barnett1915}, a rotating magnetic object is magnetized due to the coupling of angular velocity and spin. Similarly, the vorticity of a fluid couples to spin. Such effects are studied, for example, in nuclear physics \cite{Kharzeev2016} or in the context of spin hydrodynamic generation \cite{takahashi2016spin, Matsuo2017, Matsuo2017a, doornenbal2019spin}. Concerning the latter, spin currents brought about by the vorticity of a confined fluid generate nonequilibrium spin voltages. These examples have in common that a vorticity of a fluid in real space is involved. In contrast, the SCV studied here `lives' in momentum space. To put the SCV into a wider context, we show that the concept of vorticity is tightly connected to extrinsic contributions to Hall effects.

Within semiclassical Boltzmann transport theory (e.\,g., Ref.~\onlinecite{Popescu2018}) the extrinsic skew scattering contribution to the AHE is given by 
\begin{align}
    \vec{\sigma}^\mathrm{skew}_\mathrm{AHE}
    & =
    \frac{1}{2V} \sum_{n \vec{k}} \vec{J}_{n \vec{k}} \times \vec{\varLambda}_{n \vec{k}} 
    \left( - \frac{\partial f_{n \vec{k}}}{\partial \varepsilon} \right),
\end{align}
for which an AHE vector is constructed similar to the MSHE vector in Eq.~\eqref{eq:MSHEvector}. For a small electric field and a linearized Boltzmann equation, the vectorial mean free path $\vec{\varLambda}_{ n \vec{k}}$ is  obtained from \cite{Popescu2018}
\begin{align}
    \vec{\varLambda}_{n \vec{k}} = \frac{1 }{ \varGamma_{n\vec{k}}} \left( \vec{v}_{n\vec{k}} + \sum_{n', \vec{k}'} P^{n \leftarrow n'}_{\vec{k} \leftarrow \vec{k}'}  \vec{\varLambda}_{n' \vec{k}'} \right). \label{eq:mean_free_path}
\end{align}
$\varGamma_{n\vec{k}} = \sum_{\vec{k}'} P^{n' \leftarrow n}_{\vec{k}' \leftarrow \vec{k}}$ is the relaxation rate and $P^{n \leftarrow n'}_{\vec{k} \leftarrow \vec{k}'}$ is the scattering rate from a state $(n', \vec{k}')$ into a state $(n, \vec{k})$. The same steps that lead to the SCV yield 
\begin{subequations}
\begin{align}
    \vec{\sigma}^\mathrm{skew}_\mathrm{AHE}
    &=
    \frac{-e}{2\hbar (2\pi)^3} \sum_{n} \oint_{\varepsilon_{n} = \varepsilon_\mathrm{F}} \hat{\vec{v}}_{n\vec{k}} \times \vec{\varLambda}_{n \vec{k}}\, \mathrm{d}S
    \\
    &=
    \frac{-e}{2\hbar(2\pi)^3} \sum_{n} \iiint_{\varepsilon_{n} \le \varepsilon_\mathrm{F}} \vec{\nabla}_{\vec{k}} \times \vec{\varLambda}_{n \vec{k}} \, \mathrm{d}^3 k
\end{align}
\end{subequations}
for zero temperature and $\vec{J}_{n \vec{k}} = -e \vec{v}_{n \vec{k}}$. $\vec{\nabla}_{\vec{k}} \times \vec{\varLambda}_{n \vec{k}}$ is the vorticity of the mean free path ($\vec{\varLambda}$ vorticity for short). This means that a scattering process contributes to the AHE if it causes a vorticity in the mean free path; the latter is brought about by the scattering-in terms [sum in  Eq.~\eqref{eq:mean_free_path}]. To capture the skew scattering contributions to the AHE, the scattering-in terms have to be taken into account, since one finds $\vec{\vec{\sigma}}^\mathrm{skew}_\mathrm{AHE} = \vec{0}$ if these terms are neglected (e.\,g.\ in relaxation time approximation $\vec{\varLambda}_{n \vec{k}} = \vec{v}_{n \vec{k}} / \varGamma_{n \vec{k}}$). This reasoning complies with established results for the AHE \cite{Nagaosa2010}.

A skewness of the scattering is not necessary for a nonzero SCV, for the latter may be nonzero even in case of a constant relaxation rate $\varGamma_{n \vec{k}} = \varGamma$ (this is the case considered so far). If the relaxation rate depends on momentum one can still write the MSH conductivity in the form of Eqs.~\eqref{eq:MSHE_cond_vorticity} and \eqref{eq:vorticity-definition} but with a renormalized SCV
\begin{align}
    \frac{\vec{\omega}^\gamma_{n \vec{k}}}{\varGamma} & \to \tilde{\vec{\omega}}^\gamma_{n \vec{k}}
    = \frac{1}{\varGamma_{n \vec{k}}} \vec{\omega}^\gamma_{n \vec{k}} -  \vec{J}^\gamma_{n \vec{k}} \times \vec{\nabla}_{\vec{k}} \frac{1}{\varGamma_{n \vec{k}}}.
    \label{eq:correctedSCV_RTA}
\end{align}
The original SCV $\vec{\omega}^\gamma_{n \vec{k}}$ can hence be considered the backbone of the MSHE, on top of which come corrections from skew scattering, side jump or a momentum-dependent relaxation time. Future work may address the MSHE within a quantum kinetic approach, thereby taking into account the electrons' SU($2$) nature and spin-dependent scattering.

In order to avoid the ill definition of spin current, the authors of Ref.~\onlinecite{Kimata2019} considered the MSHE in terms of spin-accumulation rather than of spin-current responses. Such a reasoning fits to present experiments in which spin accumulations rather than spin currents are measured. Nonetheless, the observed spin accumulations may arise from two contributions: a local production (as for the Edelstein effect \cite{aronov1989nuclear, Edelstein1990}) and a transport of spins from the bulk toward the edges of the sample.

Compared to the spin current operator $\vec{J}^\gamma$, the velocity does not appear in the time-reversal odd spin operator $s^\gamma$. Replacing $\vec{J}^\gamma$ by $s^\gamma$ in the Kubo formula implies then that the time-reversal odd and even parts change roles; consequently, spin accumulations brought about by the MSHE appear in the intrinsic part \cite{Kimata2019} and stay finite for $\varGamma \to 0$. The latter finding is to be contrasted with the present theory which predicts a divergence of the bulk spin current in this limit [Eq.~\eqref{eq:spin_cond_odd_intra}]. This variance in one and the same limit suggests that the two underlying mechanisms are fundamentally distinct. To disentangle their relative contribution we propose that future experiments may clarify the role of relaxation processes when taking the clean limit.

\section{Concluding remarks}
\label{sec:conclusion}
We identified spin current vortices in the Fermi sea as origin of the MSHE\@. Spin current whirls in reciprocal space provide not only a vivid interpretation of the MSHE but also  corroborate that the MSHE has a bulk contribution. Future investigations in which the importance of the bulk and the interfacial contributions is considered could tell how to maximize the MSHE signal. It goes without saying that, due to Onsager's reciprocity relation \cite{Onsager1931}, the SCV also covers an inverse MSHE, that is a transverse charge current caused by a spin bias.

Having identified all magnetic Laue groups that allow for spin current vortices, we demonstrated that any ferromagnet potentially features an MSHE; furthermore, antiferromagnets whose MLG permits a magnetic toroidal quadrupole exhibit an MSHE as well. Two pyrochlore models served as examples (Sec.~\ref{sec:minimal-req}) with magnetic textures exhibiting the multipole associated with the MSH conductivities. To issue a caveat, we note that compatibility with a magnetic multipole does not necessitate the presence of the multipole. For example, a completely compensated antiferromagnetic texture may still exhibit symmetries that permit a magnetization, an observation that was appreciated in the context of the AHE for both collinear \cite{Smejkal2019CrystalHall} as well as noncollinear antiferromagnets \cite{ChenDonald2014, kubler2014non, Suzuki2017}. To name two examples: the kagome magnet discussed in Sec.~\ref{sec:MSNE} admits of a magnetization without exhibiting a net moment, and the magnetic warped Rashba model in Sec.~\ref{sec:MSHEinRashba} admits of magnetic toroidal quadrupoles although the magnetic texture is collinear.

Besides three-dimensional materials, in-plane magnetized Rashba 2DEGs with warping provide a playground for investigating a MSHE\@. A feature they have in common with Mn$_3$Sn is the option to manipulate out-of-plane polarized spin currents by rotation of the in-plane magnetic texture. The magnetization provides thus an external means to engineer spin accumulations.

Turning to magnons and replacing the electric field by a temperature gradient, our approach supports that a MSNE is expected but awaits experimental detection. The magnonic MSNE extends the family of magnonic pendants of electronic transport phenomena \cite{Mook2018}, its potential for energy harvesting and nonelectronic spin transport remains to be investigated. A candidate material for a proof-of-principle is the ferromagnetic pyrochlore Lu$_2$V$_2$O$_7$. It realizes the magnonic version of the ferromagnetic pyrochlore model of Sec.~\ref{sec:minimal-req} \cite{Elhajal2005} and is known for a thermal Hall effect \cite{Onose2010, Ideue12} and for Weyl magnons \cite{Mook2016}. For a magnetization and temperature gradient along the $[001]$ direction, we expect magnon-mediated accumulations of magnetic moments, in analogy to the spin accumulations shown in Fig.~\ref{fig:MSHEFerro}.

\acknowledgements
This work is supported by CRC/TRR $227$ of Deutsche Forschungsgemeinschaft (DFG).

\appendix

\section{Linear spin wave theory}
\label{app:LSWA}
We provide some background information on the results presented in Sec.~\ref{sec:MSNE}.
 
The directions $\hat{\vec{z}}_i$ of the spins in the classical ground state define a local coordinate system $\{ \hat{\vec{x}}_i,\hat{\vec{y}}_i,\hat{\vec{z}}_i \}$. After a (truncated) Holstein-Primakoff transformation \cite{Holstein1940} 
\begin{align}
    \frac{\vec{S}_i}{\hbar} \approx
        \sqrt{\frac{S}{2}} \left[ \left( \psi_i + \psi_i^\dagger \right) \hat{\vec{x}}_i 
        - \mathrm{i} \left( \psi_i - \psi_i^\dagger \right) \hat{\vec{y}}_i \right] 
        + \left( S - \psi^\dagger_i \psi_i \right) \hat{\vec{z}}_i,
\end{align}
from spin operators to bosonic creation and annihilation operators ($\psi^\dagger_i$ and $\psi_i$), the bilinear Hamiltonian reads
\begin{align}
    H_2 = \frac{1}{2} \sum_{\vec{k}} \adj{\vec{\varPsi}}_{\vec{k}} \underline{H}_{\vec{k}} \vec{\varPsi}_{\vec{k}}
\end{align}
after a Fourier transformation. The vector
\begin{align}
    \adj{\vec{\varPsi}}_{\vec{k}} = \qty( \adj{\psi}_{1, \vec{k}}, \adj{\psi}_{2, \vec{k}}, \adj{\psi}_{3, \vec{k}}, \psi_{1, -\vec{k}}, \psi_{2, -\vec{k}}, \psi_{3, -\vec{k}} )
\end{align}
comprises the Fourier transformed bosonic operators $\psi^{(\dagger)}_{n, \vec{k}}$ ($n=1, 2, 3$ labels the basis atoms).
The linear spin wave kernel
\begin{align}
    \underline{H}_{\vec{k}} & = \frac{S}{2}
        \begin{pmatrix}
            \underline{A}_{\vec{k}} & \underline{B}_{\vec{k}}\\
            \underline{B}^\dagger_{\vec{k}} & \underline{A}^\ast_{\vec{k}}
        \end{pmatrix}
\end{align}
is built from the submatrices
\begin{align}
    \underline{A}_{\vec{k}} & = 
        \begin{pmatrix}
            4 \left(\sqrt{3} D_z + J\right) &
            q_1 c_1 &
            q_2 c_2 \\
            \conj{q_1} c_1 &
            4 \left(\sqrt{3} D_z + J\right) &
            q_3 c_3 \\
            \conj{q_2} c_2 &
            \conj{q_3} c_3 &
            4 \left(\sqrt{3} D_z + J\right)
        \end{pmatrix}
\end{align}
and
\begin{align}
    \underline{B}_{\vec{k}} & =
        \begin{pmatrix}
            0 &
            q_4 c_1 &
            q_5 c_2\\
            q_4 c_1 &
            0 &
            q_6 c_3\\
            q_5 c_2 &
            q_6 c_3 &
            0
        \end{pmatrix},
\end{align}
with the $\vec{k}$-dependent cosines
\begin{subequations}
\begin{align}
    c_1 & = \cos \qty(a k_y),\\
    c_2 & = \cos \left (\frac{a \left(\sqrt{3} k_{x} - k_{y}\right)}{2} \right ),\\
    c_3 & = \cos \left (\frac{a \left(\sqrt{3} k_{x} + k_{y}\right)}{2} \right ).
\end{align}
\end{subequations}
Since the classical ground state is (accidentally) degenerate---the spins can be rigidly rotated within the $xy$ plane without energy cost---the $q_{i}$ ($i = 1, \ldots, 6$) depend on the chosen ground state.

Here, we consider two textures. The first texture, shown in Fig~\ref{fig:nvc}(a), is a representative of the MLG $2'/m'$; its $q_i$ read
\begin{subequations}
    \begin{align}
        q_1 & = - \sqrt{3} D_z + 2 \iu D_\parallel + J,\\
        q_2 & = - \sqrt{3} D_z + \iu D_\parallel + J, \\
        q_3 & = - \sqrt{3} D_z - \iu D_\parallel + J,\\
        q_4 & = - \left(\sqrt{3} D_z + 3 J\right),\\
        q_5 & = - \left(\sqrt{3} D_z + 3 \iu D_\parallel + 3 J\right),\\
        q_6 & = - \sqrt{3} D_z + 3 \iu D_\parallel - 3 J.
    \end{align}
\end{subequations}
The second, rotated texture, shown in Fig~\ref{fig:nvc}(d), belongs to the MLG $2/m$, 
\begin{subequations}
    \begin{align}
        q_1 & = - \sqrt{3} D_z + J,\\
        q_2 & = - \sqrt{3} D_z - \sqrt{3} \iu D_\parallel + J,\\
        q_3 & = - \sqrt{3} D_z - \sqrt{3} \iu D_\parallel + J,\\
        q_4 & = - \sqrt{3} D_z + 2 \sqrt{3} \iu D_\parallel - 3 J,\\
        q_5 & = - \left(\sqrt{3} D_z + \sqrt{3} \iu D_\parallel + 3 J \right),\\
        q_6 & = q_5.
    \end{align}
\end{subequations}

Next, we diagonalize the bilinear Hamiltonian
\begin{align}
    H_2 = \frac{1}{2} \sum_{\vec{k}} \adj{\vec{\varPhi}}_{\vec{k}} \underline{\mathcal{E}}_{\vec{k}} \vec{\varPhi}_{\vec{k}}.
\end{align}
$\underline{\mathcal{E}}_{\vec{k}} = \diag (\varepsilon_{1, \vec{k}}, \varepsilon_{2, \vec{k}}, \varepsilon_{3, \vec{k}}, \varepsilon_{1, -\vec{k}}, \varepsilon_{2, -\vec{k}}, \varepsilon_{3, -\vec{k}})$
contains the eigenvalues, and $\vec{\varPhi}_{\vec{k}} = \qty(\varPhi_{1, \vec{k}}, \varPhi_{2, \vec{k}}, \varPhi_{3, \vec{k}}, \adj{\varPhi}_{1, -\vec{k}}, \adj{\varPhi}_{2, -\vec{k}}, \adj{\varPhi}_{3, -\vec{k}})$ is a linear combination of the old bosonic operators,
\begin{align}
    \adj{\vec{\varPhi}}_{\vec{k}} & =  \adj{\vec{\varPsi}}_{\vec{k}} \underline{T}^\dagger_{\vec{k}}.
\end{align}
$\underline{T}_{\vec{k}}$ diagonalizes $\underline{H}_{\vec{k}}$ and retains the bosonic commutation rules,
\begin{align}
    \underline{T}^\dagger_{\vec{k}} \, \underline{\varSigma} \, \underline{T}_{\vec{k}} = \underline{\varSigma}, \quad \underline{\varSigma} = \mathrm{diag}(1,1,1,-1,-1,-1).
\end{align}
This procedure follows Ref.~\onlinecite{colpa1978diagonalization}.

The expectation value of the magnonic spin current of the $n$-th band is defined as \cite{Mook2019SSESNE}
\begin{align}
    J^\gamma_{n \vec{k},\mu} = \frac{1}{2} \left. \underline{T}^\dagger_{\vec{k}}  \left( \underline{v}_{\vec{k},\mu} \, \underline{\varSigma} \, \underline{s}^\gamma +  \underline{s}^\gamma \, \underline{\varSigma} \, \underline{v}_{\vec{k},\mu} \right) \underline{T}_{\vec{k}} \right|_{n,n},
\end{align}
in which $\underline{v}_{\vec{k},\mu} = \hbar^{-1} \partial \underline{H}_{\vec{k}} / \partial k_\mu$ is the velocity operator and $\underline{s}^\gamma = \mathrm{diag}(z_1^\gamma, z_2^\gamma, z_3^\gamma, z_1^\gamma, z_2^\gamma, z_3^\gamma)$ contains the $\gamma = x, y, z$ coordinate of the local spin directions for the ground state. The magnonic SCV $\vec{\nabla}_{\vec{k}} \times \vec{J}^\gamma_{n \vec{k}}$ is computed numerically; cf.\ Fig.~\ref{fig:nvc}.

\bibliography{short,newrefs}

\end{document}